\documentclass[]{article}

\usepackage{amsfonts, amsmath, amssymb}
\usepackage{fancyhdr}
\usepackage{graphicx}
\usepackage[left = 3cm, right = 3cm]{geometry}
\usepackage{float}
\usepackage{hyperref}
\usepackage[nottoc, notlot, notlof]{tocbibind}
\usepackage[utf8]{inputenc}
\usepackage[T1]{fontenc}
\usepackage[english]{babel}
\usepackage{lmodern}
\usepackage[round]{natbib}
\usepackage{subcaption}
\usepackage{float}
\usepackage{verbatim}
\usepackage{authblk}

\title{Dissecting the explanatory power of ESG features on equity returns by sector, capitalization, and year with interpretable machine learning}

\author[1,2]{J\'er\'emi Assael}
\author[2]{Laurent Carlier}
\author[1]{Damien Challet}

\affil[1]{Chair of Quantitative Finance, MICS Laboratory, CentraleSupélec, Université Paris-Saclay, Gif-sur-Yvette, France}
\affil[2]{BNP Paribas Corporate \& Institutional Banking, Global Markets Data \& Artificial Intelligence Lab, Paris, France}

\date{}

\begin{document}

\maketitle

\begin{abstract}
We systematically investigate the links between price returns and Environment, Social and Governance (ESG) scores in the European equity market. Using interpretable machine learning, we examine whether ESG scores can explain the part of price returns not accounted for by classic equity factors, especially the market one. We propose a cross-validation scheme with random company-wise validation to mitigate the relative initial lack of quantity and quality of ESG data, which allows us to use most of the latest and best data to both train and validate our models. Gradient boosting models successfully explain the part of annual price returns not accounted for by the market factor. We check with benchmark features that ESG data explain significantly better price returns than basic fundamental features alone. The most relevant ESG score encodes controversies. Finally, we find the opposite effects of better ESG scores on the price returns of small and large capitalization companies: better ESG scores are generally associated with larger price returns for the latter and reversely for the former.
\end{abstract}



\section{Introduction}  
\label{sec:introduction}

Investing according to how well companies do with respect to their Environmental, Social and Governance scores has become very appealing to a growing number of investors. Beyond moral criteria, such kinds of investments may increase the value of high-ESG-scoring companies, which will attract even the non-ESG-minded investor, thereby starting a virtuous circle both for the investors and for the beneficiaries of high ESG scores. It may also lead to successful impact investing whereby an investor generates positive environmental or societal impact while targeting a specific level of return \citep{townsend2020sri, grim2020esg}. 

From a quantitative point of view, ESG scores raise the question of their information content: do these scores contain some signal to estimate a company's fundamental or market information? Restricting themselves to the study of the explanatory and predictive power of ESG scores regarding financial performance, \cite{friede2015esg} aggregate the results of more than 2200 studies: 90\% of them show a non-negative relationship between ESG and corporate financial performance measures, a majority displaying a positive relationship. However, more recently,  \cite{cornell2020valuing}, \cite{breedt2019esg}, and \cite{margot2021esg} reached less clear-cut conclusions. 

The confusion surrounding this question  is mostly caused by the nature of ESG data: (i) they are quite sparse before 2015, as the interest in even computing such scores is quite recent; (ii) they are usually updated yearly; (iii) the way they are computed often changes as a function of time and may depend on the way companies disclose data; (iv) human subjectivity may be involved to a large extent in the computation of the scores, according to the methodology chosen by a given data provider. Findings are therefore inevitably data-vendor dependent. While data consistency and quality can only be solved at the data provider level, points (i) and (ii) require a tailored approach.

Here, we argue that settling this issue requires a globally robust and consistent methodology. We discuss how to solve each of the two remaining problems listed above and propose a methodology that combines a novel cross-validation procedure for time series with increasingly reliable data, explainable machine learning, and multiple hypotheses testing. Although we focus on explaining company price returns with ESG scores, this methodology can be easily adapted and extended to different settings (e.g., prediction). 

Another crucial ingredient of our approach is to focus on the simplest possible question. Instead of performing sophisticated regressions, we aim to explain the sign of excess price returns. From an information-theoretic point of view, this means that we focus on a single bit of information (the sign) instead of many bits (full value), which yields significant and robust results that  can then be interpreted as a function of market capitalization, industrial sector and country.

Our contributions are as follows:

\begin{enumerate}
    \item We focus on the sign of either excess returns (main text) and returns discounted by Fama French factors (appendix) and use state-of-the-art sign prediction machine learning models;
    \item We propose a company-wise cross-validation scheme that makes it possible to train and validate models with the most recent (and thus most reliable) data; from this validation scheme, we keep the models with the five best validation scores;
    \item We show that the fitted models explain well the signs of excess returns in test periods that are not used to calibrate these models. We also show that models trained with ESG scores increasingly outperform models trained with fundamental data only;
    \item Finally, we show how each individual ESG score contributes to the overall performance of our algorithm and the evolution of their explanatory power as a function of time. We propose a new way to build a so-called materiality matrix based on the interpretability of the chosen machine learning models, showing that the importance of ESG scores depends on both the industrial sectors and market capitalization.
\end{enumerate}

In the remainder of this study, the terms ESG scores and ESG features are used interchangeably.

\section{Literature review and uncertainties}

\subsection{Asset selection, investment strategies, and portfolios}
According to \cite{chen2020integrated}, ESG integration into investment strategies mainly consists in integrating the investors’ values into their own strategies. The scientific literature describes three main ways to achieve it: filtering companies based on their ESG scores, directly looking for alpha in ESG data, or measuring ESG impact on other risk factors.

ESG scores can offer a systematic approach to screen out controversial industries, commonly referred to as ``sin industries'',  including but not limited to tobacco, alcohol, pornography, weapons, etc. For example, some studies advocate for selecting companies with ESG scores surpassing specific thresholds \citep{schofield2019whattolook}. While this method yields good portfolios ESG-wise, \cite{alessandrini2020esg} argue that this may lead to underperforming portfolios due to the reduction in the investment universe and the potentially higher returns generated by ``sin industries'' because of their very exclusion. 

\cite{chen2020integrated} propose a Markowitz-like optimization method by defining an ESG-compatible efficient frontier. Similarly, \cite{hilario2020tri} add  a third term to the mean-variance cost function, the portfolio exposure to carbon risk, and use a genetic algorithm to solve this three-criterion optimization problem. This method is equivalent to optimizing ESG criteria under the constraint of specific risk and returns levels \citep{schofield2019whattolook}, who also note that the resulting portfolio can have a good global ESG score while containing assets with bad ones.

Finally, \cite{alessandrini2020esg} elaborate on ``smart beta'' strategies, in which investors  build portfolios whose assets are not weighted according to their market capitalization but rather to their exposure to some specific risk factors. \cite{bacon2015smart} explain that integrating ESG into investment strategies can be simply achieved by tilting the asset weights according to their ESG scores while controlling the portfolio exposure to other risk factors. This procedure raises the question of whether ESG is a new risk factor or if optimizing  ESG scores amounts to exposing the portfolio to well-known ones. It is indeed a crucial point to explore  when attempting to improve portfolio performance with ESG scores
\citep{anson2020sustainability}: instead of trying to obtain a premium by finding a suitable ESG factor, it is more judicious to understand the impact of ESG data on the exposure to well-known risk factors.

\subsection{ESG scores: risk and returns}
Reaching a consensus on the nature of the links between ESG and returns is hard. \cite{friede2015esg}  aggregate more than 2000 studies on the topic: 41\% did not find any ESG impact on returns, 48\% found these impacts to be positive and 9\% negative. \cite{alessandrini2020esg} and \cite{anson2020sustainability} stress the fact that filtering a portfolio on ESG scores leads to  improved durability of the investment but does not yield a positive alpha. However, they did not find any proof of negative alpha either; thus, there may be no added value in integrating ESG data into portfolio construction from an alpha point of view.

\cite{plagge2020have} find no statistically significant under- or over-performance  of different equity funds specialized in ESG investing and argue that since the ESG scores are not of economic nature, they should not have any impact on the portfolios and that any information contained in ESG data should  already be contained in other risk factors. However, \cite{lee2022proposing} find, using machine learning methods, that ESG granular data of considered equity funds provide information on the annual financial performance of these funds. 

This lack of consensus on the links between ESG and returns may be due to the use of different assessment methodologies (including ESG scores from different data providers) or wrong use of the ESG scores \citep{anson2020sustainability}. Indeed, \cite{margot2021esg} show that because ESG data have a very low signal-to-noise ratio, using aggregated ESG scores leads to a high loss of information. It is then necessary to use more granular scores to obtain more meaningful results. Moreover, they emphasize that the links between ESG and returns are highly dependent on the considered business industry and region. \cite{cappucci2018esg} finds that ESG scores lack information on asset price returns and that a better indicator of returns is the progress made by companies in the different ESG sub-fields.

Only a few papers are devoted to the relationship between ESG scores and risk. \cite{guo2020esg2risk} train a deep learning model to predict a company’s volatility using ESG news. \cite{chen2020integrated} show that focusing on ESG investments can reduce the risk of underperformance as companies with good ESG scores can be less exposed to both systemic and idiosyncratic risks.

\subsubsection{Risk factors}
Many studies, such as \cite{renshaw2018esg}, find that ESG scores and well-known risk factors, such as size, are partially redundant. \cite{anson2020sustainability} and \cite{kanqui2019whyusingesg} study the variation of portfolio exposure to well-known risk factors when one integrates ESG data in portfolio construction: the impact varies according to geographical regions, which reduces the  significance of global studies. Similarly, \cite{alessandrini2020esg} explain that the discrepancies in ESG portfolio performances in different regions and industries can be attributed to different exposures to risk factors. Furthermore, \cite{breedt2019esg} argue that most of the financial performance of a portfolio can be explained by well-known factors and that the residuals cannot be explained by any other factors. For \cite{breedt2019esg}, the environmental and social aspects of ESG are noise, and the governance part is strongly correlated to the quality factor; however, enriching ESG data with other types of information, or  preprocessing it, can bring added value. In the same vein, \cite{bacon2015smart}  decorrelate the ESG scores from the other risk factors before integrating them into strategies and  are able to obtain added value from ESG scores. 

\subsubsection{Materiality of ESG data}
For a better ESG integration, it is important to understand which ESG features are the most material, i.e.,  have the largest impact on the financial performance of a company. According to \cite{anson2020sustainability} and \cite{margot2021esg}, materiality is highly dependent on the chosen asset class, region and industry. \cite{bacon2015smart}  build a materiality matrix using the LASSO method \citep{tibshirani1996regression}. Their matrix is specific to an industry and shows the magnitude of the impact of a specific ESG feature on a company's financial performance versus the probability of this feature having an impact.

\subsubsection{Temporality of ESG data}
\cite{alessandrini2020esg} warn that their results were obtained in a period when a large amount of money was poured into ESG funds, which could have increased their respective performance. \cite{margot2021esg} also reinforce that their study was realized between 2009 and 2018 during a period when the market was particularly bullish, which may affect the overall strength of ESG-based funds. 
For \cite{mortier2019alpha}, the impact of ESG scores differs not only by region and by industry but also according to the chosen period to test the strategy. That is why \cite{renshaw2018esg} argues that any methodology that treats the historical ESG data in the same way for every period is likely not relevant. A solution is to use back-testing on several time periods, with several universes, to validate the results \citep{anson2020sustainability}. Finally, \cite{margot2021esg} and \cite{plagge2020have} apply the efficient markets theory in the context of ESG investing: it is possible that investor awareness rises as a function of time and the information included in ESG data is included in the asset prices, leading to a loss of predictive power of ESG features and thus of the embedded alpha.

\section{Datasets}
	
\subsection{Financial data}

We use the following data sets:
\begin{itemize}
	\item Stock prices. We use daily close prices, adjusted for dividends and foreign exchange rates. BNP Paribas internal data sources.
	\item Market capitalization. BNP Paribas internal data sources.
	\item Fama--French market, size and value factors: these factors are taken from the online French data library \citep{dataFamaFrenchWebsite}. They are all computed according to the Fama and French methodology exposed in \cite{fama1993common}.
	\item Risk-free rate: these  data are also taken from \cite{dataFamaFrenchWebsite} and computed according to the  Fama and French method.
\end{itemize}

In addition, metadata such as the TRBC (The Refinitiv Business Classification) sector at levels 1, 2 and 3 and the country of incorporation are used and come from Refinitiv data sources.

\subsection{ESG data}

\begin{figure}
		\centering
		\includegraphics[width=.5\linewidth]{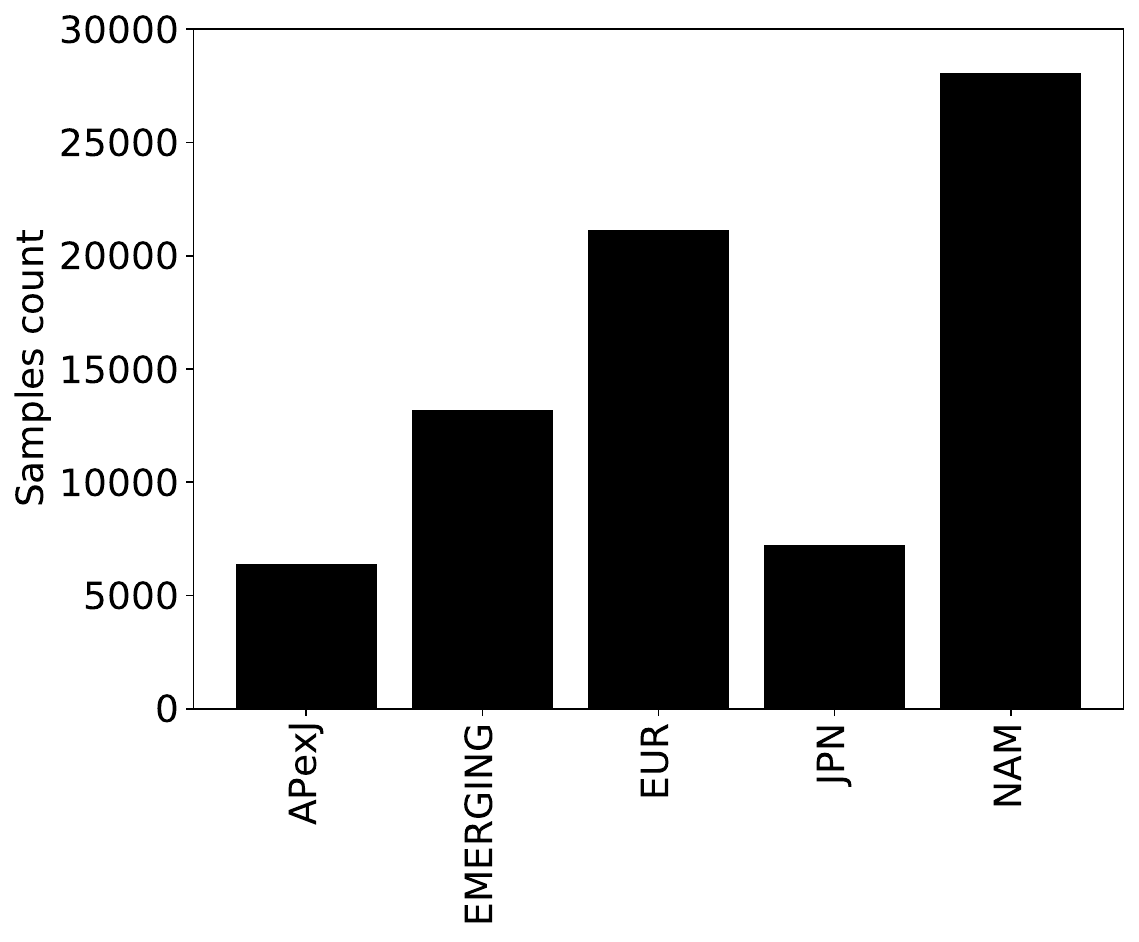}  
		\caption{Number of samples in each Fama-French region in the Refinitiv ESG dataset.}
		\label{fig:samplesperregion}
\end{figure}

\begin{figure}
		\centering
		\includegraphics[width=.5\linewidth]{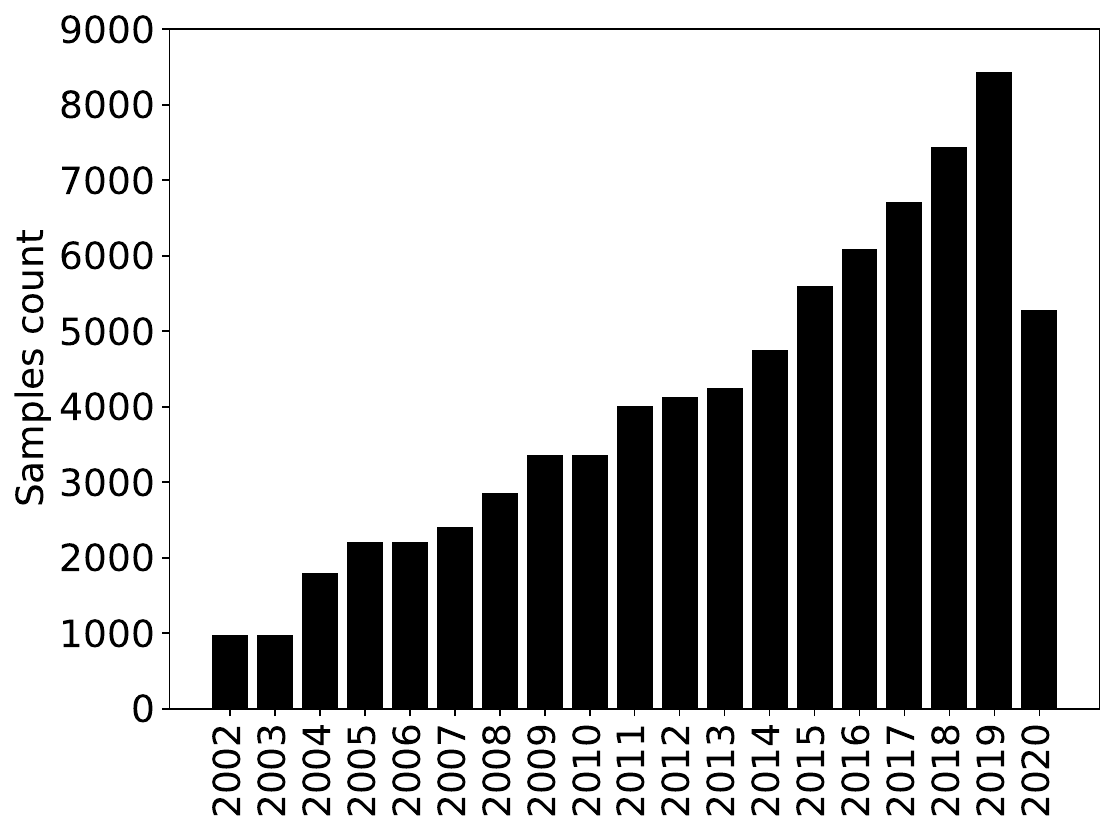}  
		\caption{Time evolution of the number of samples in the Refinitiv ESG dataset.}
		\label{fig:samplesperyear}
\end{figure}

ESG data are provided by Refinitiv. Their database alleviates some of the challenges listed above:
\begin{enumerate}
    \item The coverage of the dataset is sufficient to extract meaningful results. Figure \ref{fig:samplesperregion} shows the number of samples in the geographical regions as defined by Fama and French  \citep{dataFamaFrenchWebsite}: Europe, North America, Japan, Asia-Pacific excluding Japan and emerging countries. Refinitiv  ESG data started in 2002, and the number of samples per year increased  several-fold until 2019, as shown in Figure \ref{fig:samplesperyear}. The drop in 2020 is due to the fact that not all the ESG scores had been computed by Refinitv when we had access to the dataset (many companies had not yet published enough data).
    \item Scores are built with a well-documented methodology explained in \cite{refinitiv2021methodo}. Every ESG score ranges between 0 and 1, with 1 being the best score. In addition, the same methodology is used throughout the years, yielding consistent data.
    \item Human intervention is limited to some quality checks.
    \item Scores can be updated up to 5 years after the first publication, which is beneficial in an explanatory setting, as the data become more accurate. In a purely predictive setting, however, this adds noise and look-ahead bias as we do not have point-in-time data, i.e., we do not know the initial and intermediate ESG estimates.
\end{enumerate}

Refinitiv ESG data includes samples from different regions of the world. Each region has specific regulatory frameworks and ESG transparency rules. This is why this paper focuses on the European region and includes all the companies in the Refinitiv ESG dataset whose country of incorporation is in Europe or in a European-dependent territory.

\begin{figure}
		\centering
		\includegraphics[width=.5\linewidth]{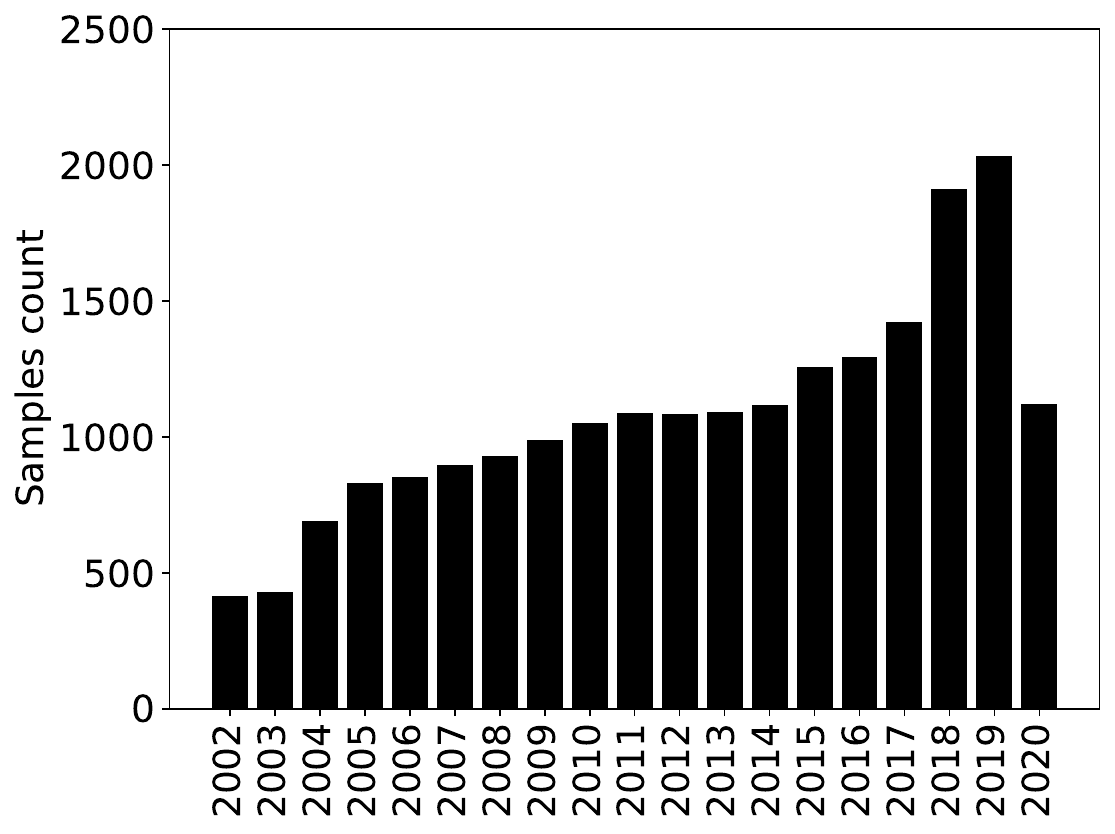}  
		\caption{Time evolution of the number of samples per year in the Refinitiv ESG dataset - Europe.}
		\label{fig:sampleseurperyeareur}
\end{figure}

\begin{figure}
		\centering
		\includegraphics[width=.5\linewidth]{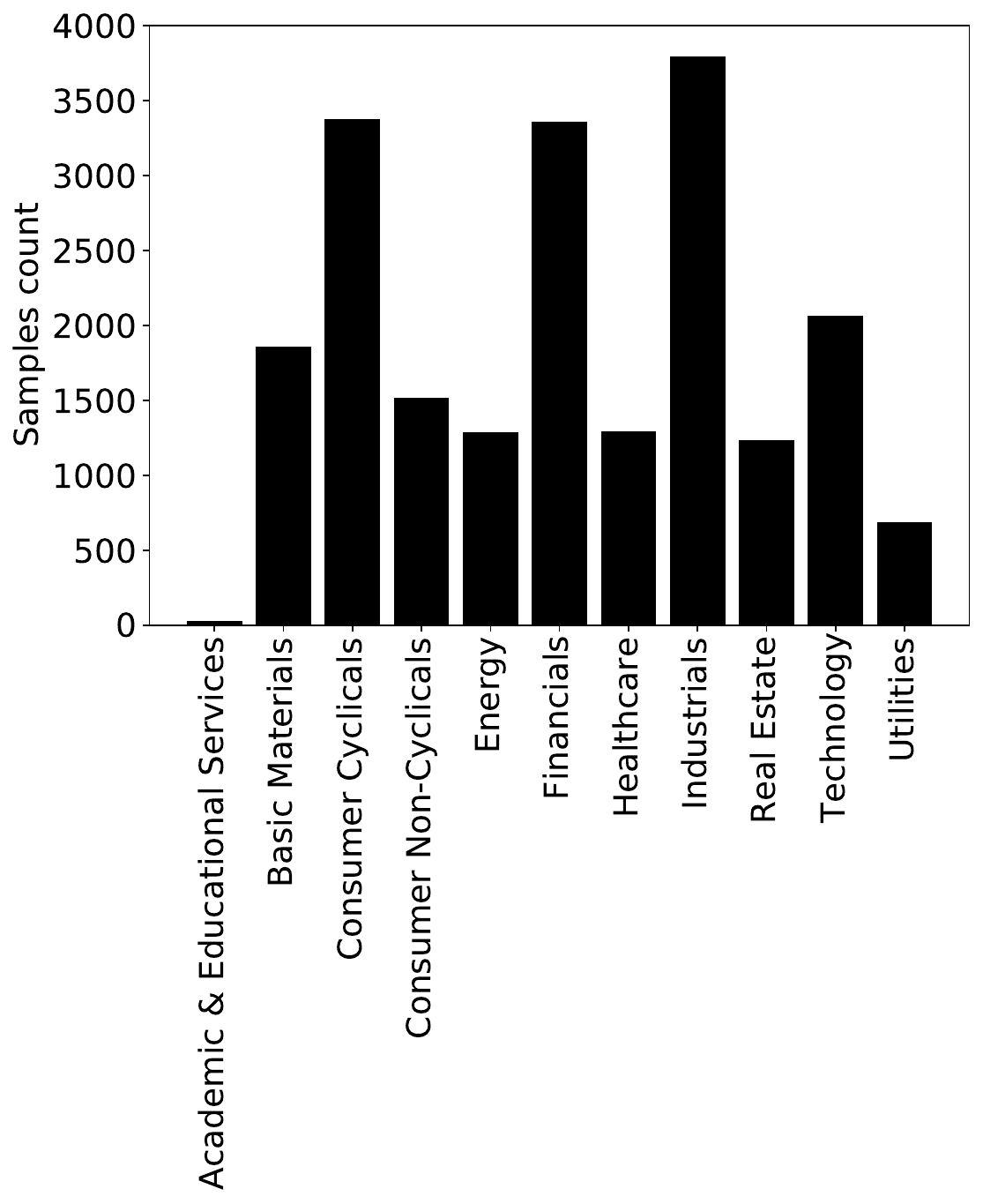}  
		\caption{Number of samples for TRBC L1 Sector in the Refinitiv ESG dataset - Europe.}   
		\label{fig:sampleseurperl1eur}
\end{figure}

The European ESG dataset contains 20,509 samples for 2429 companies uniquely identified by their ISIN. The time evolution of the number of samples per year is reported in Figure\ \ref{fig:sampleseurperyeareur}. All the sectors have enough data, with the notable exception of the Academic and Educational Services sector (see Figure\ \ref{fig:sampleseurperl1eur}).


\section{Methods}

\subsection{Problem settings}
\label{sub:problemsettings}

Our goal is to understand how and what ESG features participate in the formation of price returns. Specifically, we seek to investigate whether ESG features help capture information to explain the parts of stock returns realized at the time of the publication of the ESG data that are not accounted for by well-known equity factors, especially the market, size, and value factors. In a multi-factor model, one writes at time $t$
\begin{align}
	\label{eqn:ff3}
	r_{i,t}  =  r_{f,t} + \sum_k w_{i,k} F_{k,t} + \alpha_i + \epsilon_{i, t}
\end{align}
where $r_{i,t}$ is the return of asset $i$, $r_{f,t}$ the risk-free rate, $F_{k,t}$ the value of factor $k$ at time $t$ and $w_{i,k}$ is the factor loading; the idiosyncratic parts are $\alpha_i$, the unexplained average return, and the zero-average residuals $\epsilon_{i,t}$. In this work, we use the Capital Asset Pricing Model (CAPM) and its extension, the Fama--French 3-factor model that includes market ($r_{m}$), size Small Minus Big (SMB), and value High Minus Low (HML) factors \citep{fama1993common}. 

ESG data are neither abundant nor of constantly high quality. Directly estimating the explanatory power of ESG features on price returns by estimating the idiosyncratic part of Equation \eqref{eqn:ff3}, $\alpha_i + \epsilon_{i, t}$, is a challenging task. Therefore, we settle in this study for a less ambitious goal. Specifically, we investigate whether ESG features help explain the sign of the idiosyncratic part of price returns. Mathematically, one needs to explain
\begin{equation}\label{eq:Y}
Y_{i,t}=  \begin{cases}
    1, & \text{if $\textrm{sign}(\alpha_i+\epsilon_{i,t})=0$}.\\
    \frac{1+\textrm{sign}(\alpha_i+\epsilon_{i,t})}{2}, & \text{otherwise}.
    \end{cases}
\end{equation} 
with the candidate features. Equation \eqref{eq:Y} means that the chosen target is 0 if the sign of the idiosyncratic parts $\alpha_i$ + $\epsilon_{i,t}$ is negative, and 1 if this sign is positive or null.

This work takes a machine learning approach to this problem and treats it as a classification problem:  $Y_{i,t}$ defines two classes as it can take two values. Thus, for each possible couple $Y_{i,t}$, one has a vector of $P$ potentially explanatory factors, called features in the following. Let us relabel all the couples $(i,t)$ by the index $n\in\{1,\cdots,N\}$. The classification problem consists in explaining $Y_n$ by a vector $X_n$ with $P$ components, or equivalently, explaining the vector $Y\in\{0,1\}^{N}$ from the lines of matrix $X\in\mathbb{R}^{N \times P}$. $Y$ is called the target and $X$ the feature matrix. The problem is then to train a machine learning method to learn the mapping between the lines of $X$ and the components of vector $Y$. Once the training is complete, such a model takes  a vector of features as the input and outputs the probability that these features correspond to one class (in a two-class problem).

The state-of-the-art for these tabular data is Gradient Boosting models \citep{friedman2001greedy}, as shown, for instance, in \cite{shwartz2021tabular}. The spirit of gradient boosting consists in using a sequence of weak learners (wrong models) that iteratively correct the mistakes of the previous ones, which eventually yields a strong learner (good model). We use  decision trees here as weak learners. Different implementations of the Gradient Boosted Decision Trees method exist, e.g., XGBoost \citep{chen2016xgboost}, LightGBM \citep{ke2017lightgbm}, CatBoost \citep{prokhorenkova2017catboost}. We use  LightGBM here. 
One of the primary advantages of such methods over logistic regression is their ability to learn more generic, non-linear functional forms, which in turn yields superior performance.  While deep learning was not employed in this paper, a number of studies, such as \cite{schmitt2022deep} or \cite{shwartz2021tabular}, have shown that gradient-boosted models are at least as effective as deep neural networks for classification purposes in the context of tabular data. Moreover, gradient-boosted models are typically much faster to train than deep neural networks. Recurrent deep learning models, such as Long Short-Term Memory (LSTM) networks, suffer from similar drawbacks and require even longer training times. Furthermore, such models are not well-suited to the dataset under consideration in this study, given that some companies have only one or two years of history, resulting in very small sequence lengths.

The models are trained to minimize the cross-entropy (cost function), also known as LogLoss, defined as:
\begin{equation}
\mathcal{L} = -\frac{1}{N} \sum_{i=1}^{N} y_i \log(p_i) + (1-y_i)\log(1-p_i),
\end{equation}
where $p_i$ is the model probability that sample $i$ belongs in category 1 and $y_i\in\{0,1\}$ is the true class (which selects the suitable term of the sum for each $i$). This type of loss implicitly assumes that both true classes appear with roughly the same frequency in the training set, which is the case with 51.7\% of samples belonging to class 1 and 48.3\% to class 0.

\subsection{Training features}

The Refinitiv ESG dataset contains several levels of granularity. We choose to train our models with the 10 pillar scores described in Appendix \ref{app:10pillarscores} (Resource Use, Emissions, Innovation, Workforce, Human Rights, Community, Product Responsibility, Management, Shareholders, CSR Strategy) and the aggregated Controversy score. This level of granularity is a good compromise.

We add five non-ESG features (market capitalization, country of incorporation and TRBC sectors at levels 1, 2 and 3). These features provide the benchmark features needed to settle the question of the additional information provided by ESG features.

\subsection{Target computation}

We compute the coefficients of the regression defined in Equation \eqref{eq:Y} with monthly factors available online at \cite{dataFamaFrenchWebsite} and monthly price returns over periods of 5 civil years. For instance,  the regression coefficients used to compute the 2017 target, possibly explained by 2017 ESG features, are computed with historical data ranging from 2013 to 2017. We then compute targets over the year corresponding to the year of the publication of the ESG features: as we are in an explanatory setting, we want to explain the return of a company for a specific year using the ESG profile of this company during the same year.

\subsection{Cross-validation and hyperparameter tuning in an increasingly good data universe}
\label{sub:splittingstrat}

The usual strategy of a single data split into a causal consecutive train, validation and test data sets may not be fully appropriate for the currently available ESG features. This is because the amount of data grows from a very low baseline, both quantity- and quality-wise, which was not exploitable, to an amount that more likely is. Thus, not only are the data non-stationary but their reliability and quality keep increasing. As a consequence, the cross-validation time-splitting schemes known to work well in the context of non-stationary time series \citep{bergmeir2012use} may be improved upon.

\begin{figure}
	\centering
	\includegraphics[width=1\linewidth]{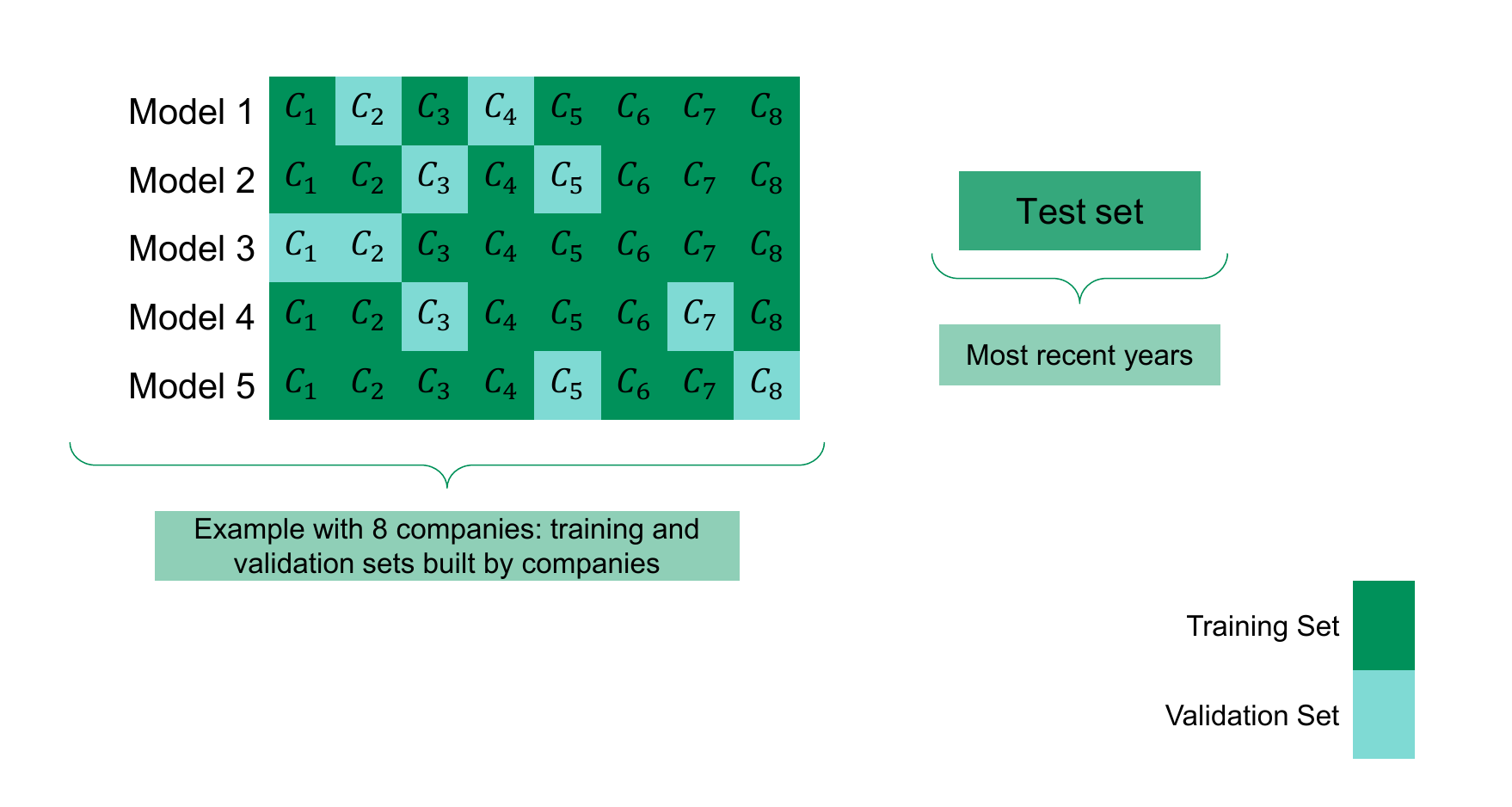}  
	\caption{Company-wise cross-validation: the validation sets consists of randomly selected companies, which allows training to account for most of the most recent data.}
	\label{fig:compstrat}
\end{figure}

For this reason, we experiment with $K$-fold company-wise cross-validation, where 75\% of companies are randomly assigned to the training set and the remaining 25\%   to the validation set (see Figure\ \ref{fig:compstrat}). In other words, there are $K$ different (train-validation) sets. For each of the $K$ train sets, we train 180 models, varying 12 hyperparameters of the LightGBM (maximum tree depth, learning rate, etc.) and pool the five best ones according to model performance in the respective validation sets. In this way, models are trained with most of the most recent (hence, more relevant) data while also being validated with the most recent and best data. If the dependencies completely change every year, this validation scheme is bound to fail. As we shall see, this is not the case. We take $K=5$.

In addition, we use expanding (train+validation)-test windows, using the last year as the test window, which allows us to perform a time-wise analysis of the performance of the models. Because data are insufficient before 2015, we have five different periods: the first test year is 2016, and the last one is 2020. We thus train and validate $K\times 5=25$  models.

For each testing period, we will compare the performance of the company-wise $5$-fold random splits with that of the standard temporal split (75\% train/25\% validation).

\section{Results}
\label{sec:results}

Here we investigate the results of the standard temporal split and the $5$-fold company-wise split for a target computed using the CAPM model, as described in Section \ref{sub:problemsettings}. Models trained using the Fama--French 3-factor model lead to less clear-cut performance; their results are relegated to Appendix \ref{app:targetstests}.

We first assess the quality of the models according to the cross-entropy loss, using their direct probability outputs. We also assess the end result, i.e., the predicted class. As it is usual, we map the output, a probability $p_i$, to classes 0 and 1 with respect to a 0.5 threshold. This allows us to compute the balanced accuracy, defined as the average of the sensitivity and the specificity. Sensitivity equals the ratio of true positives to the number of positive samples. Specificity is the ratio of the true negatives to the number of negative samples. An advantage of  balanced accuracy over classical accuracy is that balanced accuracy accounts for class imbalance in the test set. By definition, it assigns a score of 0.5 if the model did not learn anything significant.

\begin{figure}
	\begin{subfigure}{0.33\linewidth}
		\centering
		\includegraphics[width=1\linewidth]{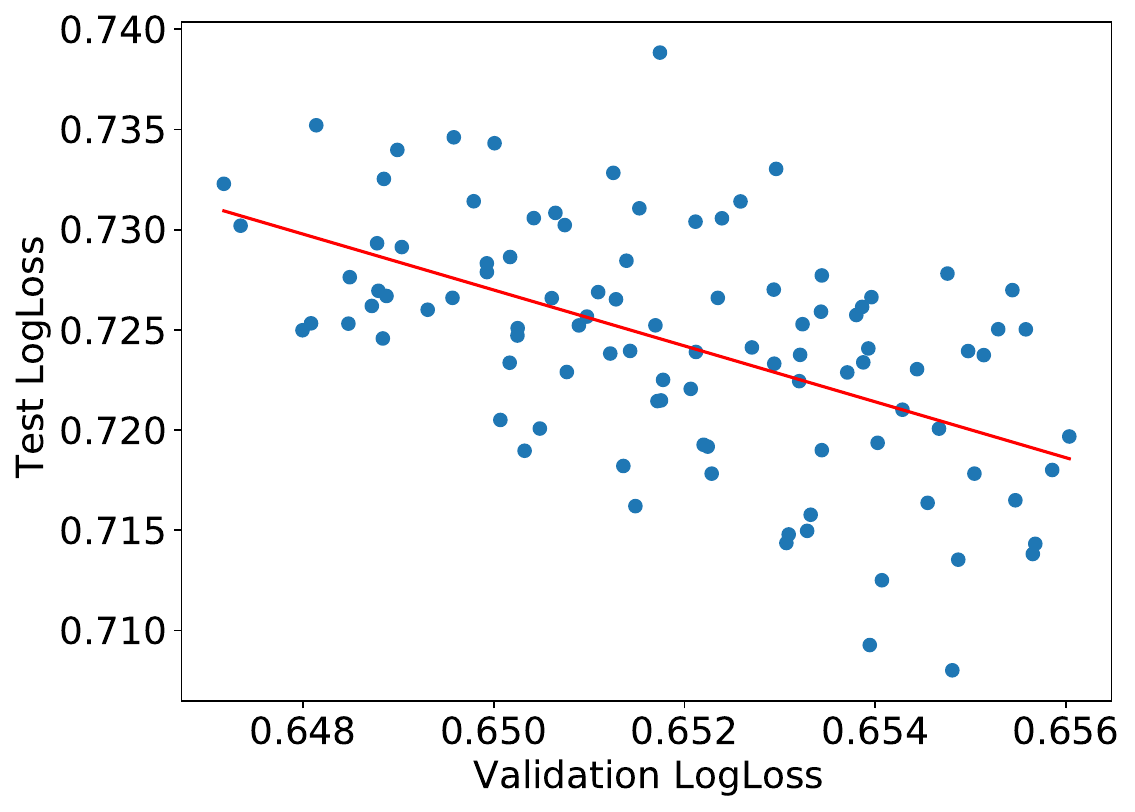} 
		\caption{2016}
		\label{fig:INST_S5_42370_-1.39_0.29}
	\end{subfigure}
	\begin{subfigure}{0.33\linewidth}
		\centering
		\includegraphics[width=1\linewidth]{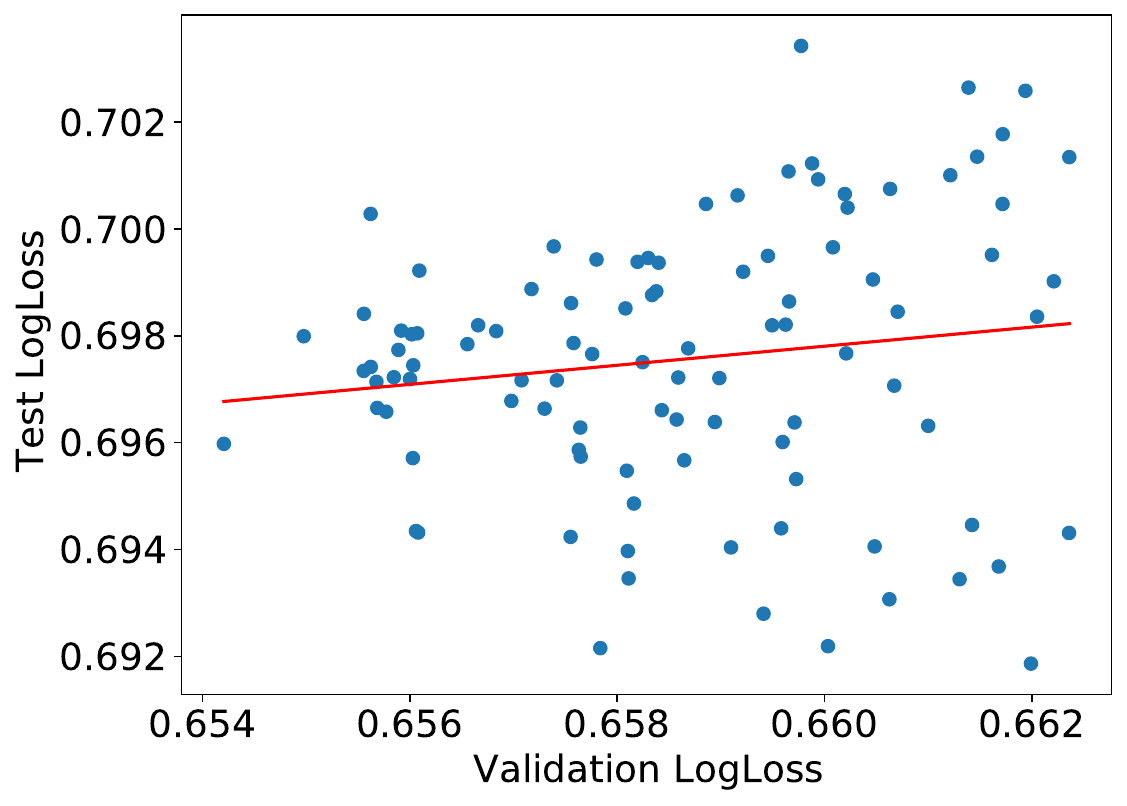}  
		\caption{2017}
		\label{fig:INST_S5_42736_0.18_0.02}
	\end{subfigure}
	\begin{subfigure}{0.33\linewidth}
		\centering
		\includegraphics[width=1\linewidth]{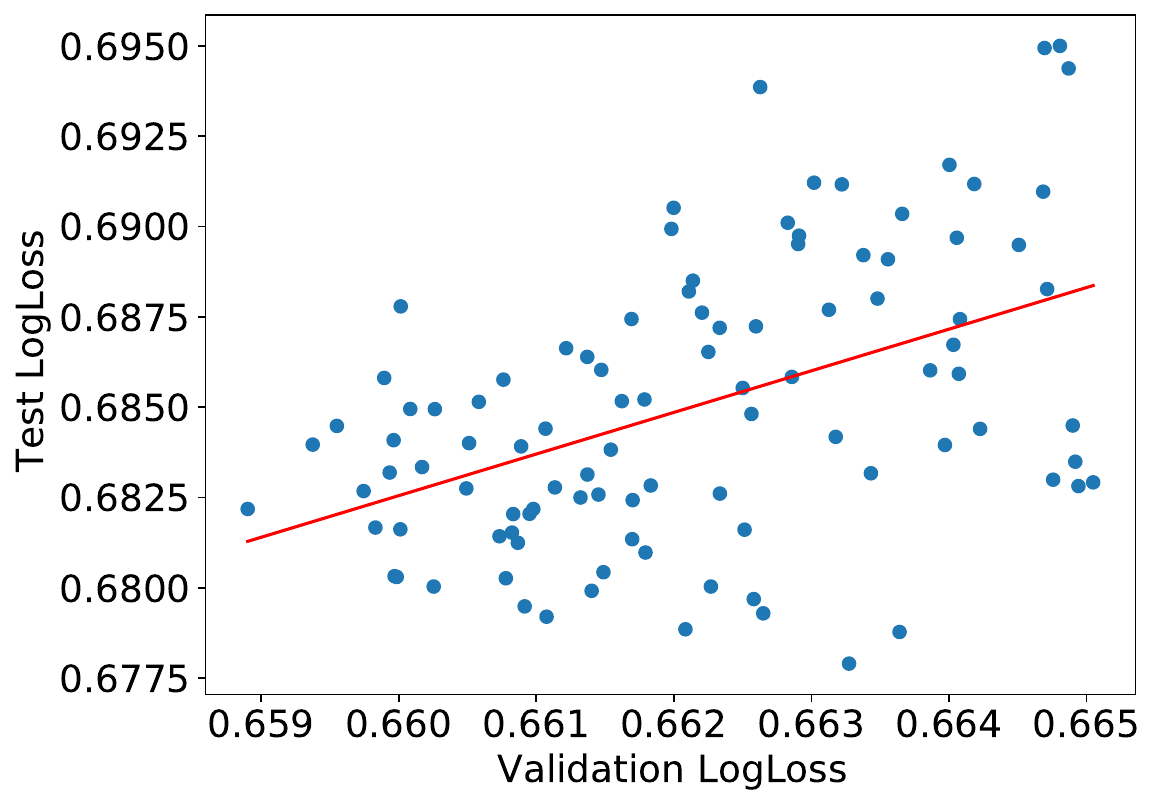}  
		\caption{2018}
		\label{fig:INST_S5_43101_1.15_0.22}
	\end{subfigure}
	\begin{subfigure}{0.33\linewidth}
		\centering
		\includegraphics[width=1\linewidth]{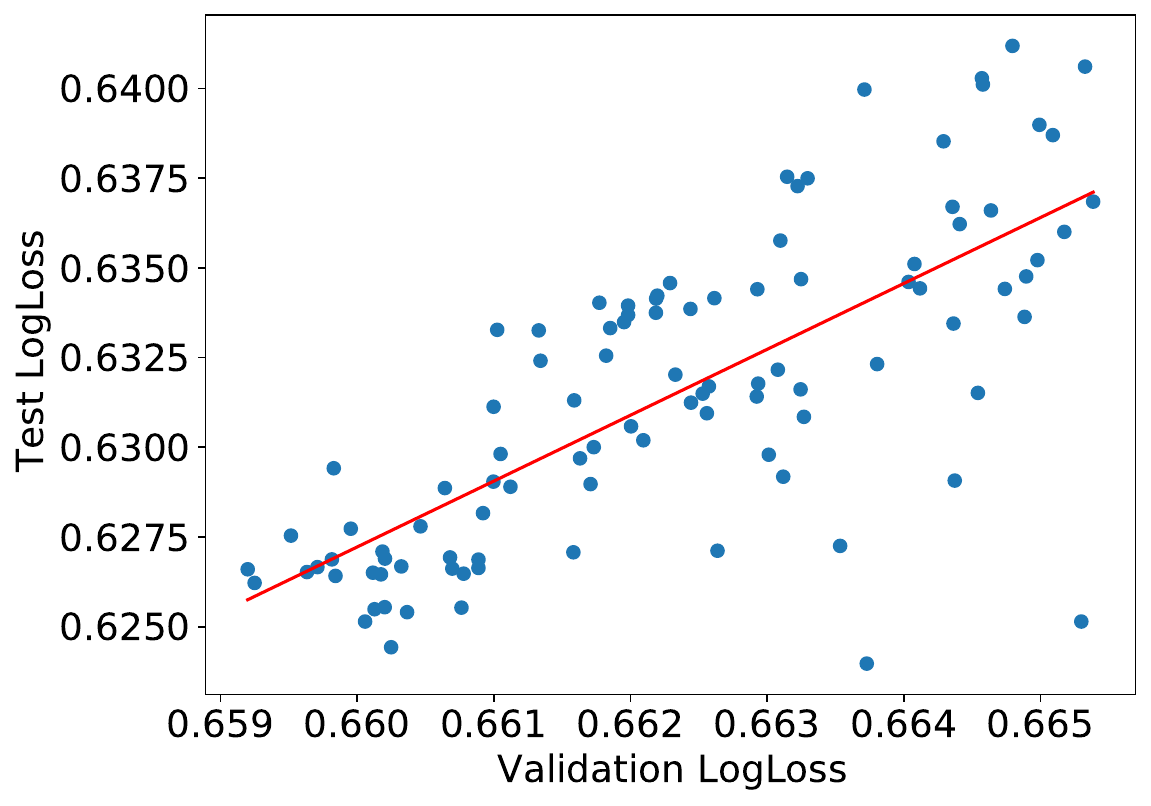}  
		\caption{2019}
		\label{fig:INST_S5_43466_1.84_0.54}
	\end{subfigure}
	\begin{subfigure}{0.33\linewidth}
		\centering
		\includegraphics[width=1\linewidth]{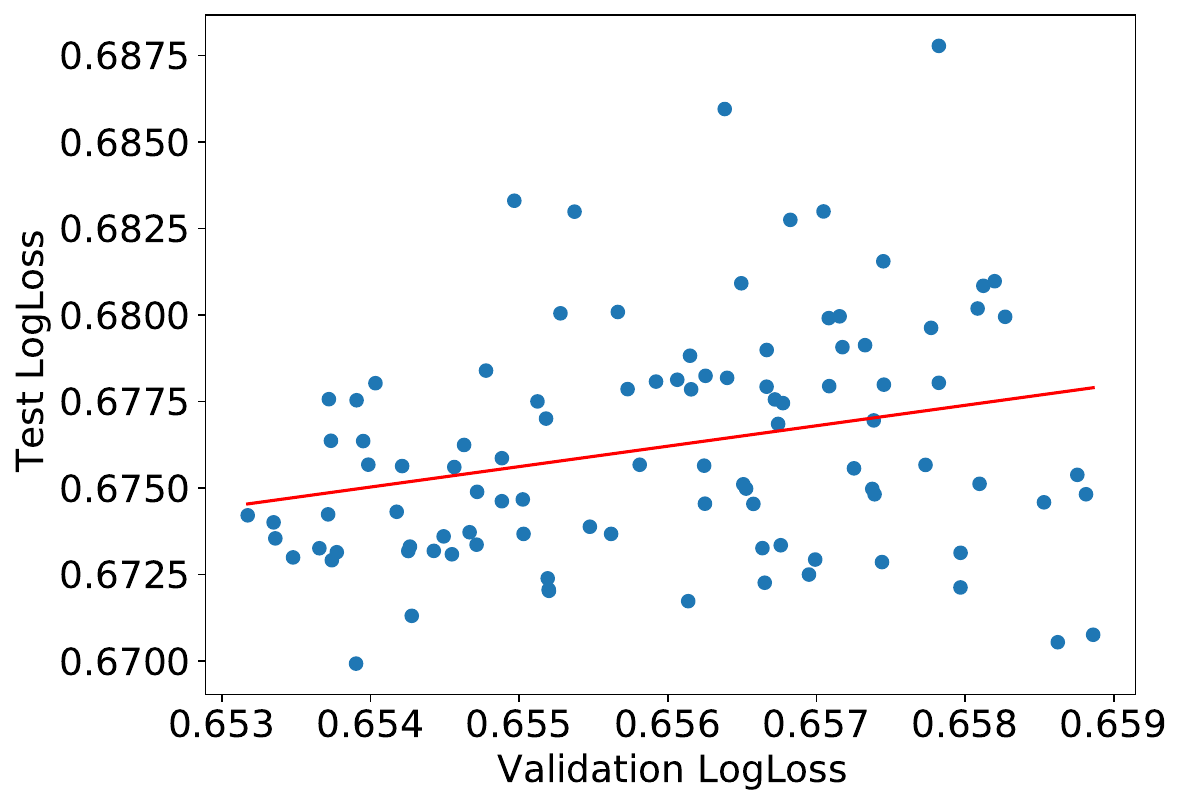}  
		\caption{2020}
		\label{fig:INST_S5_43831_0.59_0.07}
	\end{subfigure}
	\caption{Company-wise cross-validation: test set cross-entropy versus validation cross-entropy of the 100 best models of the random hyperparameters search.}
	\label{fig:refinitivresultsgraph_CAPM_INSTS5}
\end{figure}

We check that the performance of the models in the test sets bears some relationships with their performance in the validation sets. More precisely, for each (train+validation)-test period, we investigate the dependence between the cross-entropy losses in the validation and test sets, respectively noted $\mathcal{L}_m^\textrm{validation}$ and $\mathcal{L}_m^\textrm{test}$, for the best models trained during the hyperparameters random search, which makes it possible to characterize the training quality year by year. A significantly positive relationship shows that these models did learn persistent relationships, i.e., something useful. Mathematically, we assess the relationship $\mathcal{L}_m^\textrm{test}$ versus $\mathcal{L}_m^\textrm{validation}$ for each model $m$ ranking in the 100 models with the best validation cross-entropy losses for each of the five sets of (train+validation)-test sets.
Figure \ref{fig:refinitivresultsgraph_CAPM_INSTS5} displays these relationships for the company-wise cross-validation scheme and adds a linear fit. Figures of the same type for the standard time-splitting scheme can be  found in Appendix \ref{app:graphs_val_test}. Generally, both test and validation cross-entropy losses are positively correlated, except for $2016$. We believe that this comes from the fact that ESG data were of insufficient quality before that date. The year $2020$ is also special: in addition to the coronavirus crisis, the data for 2020 were obtained at the beginning of 2021 when not all companies had ESG ratings, leading to a smaller dataset and a (mostly likely) biased test set.

We compute the Pearson correlation, the $R^2$ of the linear fit, Kendall tau and its p-value for the standard temporal split and the $5$-fold company-wise split, which are reported in Table \ref{tab:validation_test_corr_table}. This latter allows us to compare the respective advantages and disadvantages of each validation strategy. All the dependence measures increase significantly from $2017$ to $2019$ for company-wise splits. The case of temporal split shows the limitations of this approach: the performance measures are roughly constant, which is consistent with the fact that adding one year of data to the train+validation dataset does not lead to much change.  A display of the relationship $\mathcal{L}_m^\textrm{test}$ versus $\mathcal{L}_m^\textrm{validation}$ for the standard temporal model can be found in Appendix \ref{app:graphs_val_test}, in Figure \ref{fig:refinitivresultsgraph_TEMP}.

\begin{table}
	\centering
	\resizebox{\textwidth}{!}{\begin{tabular}{l||c|c|c|c}
		{} & \multicolumn{4}{c}{\textbf{Company-wise $5$-fold cross-validation}}\\
		\hline
		Year &  Pearson correlation & $R^2$ &  Kendall tau & p-value of Kendall tau \\
		\hline
		2016 &      -0.54 &    0.29 &     -0.36 &     $8.0\mathrm{e}^{-8}$ \\
		2017 &      0.14 &     0.021 &     0.12 &     $6.7\mathrm{e}^{-2}$ \\
		2018 &      0.47 &     0.22 &     0.30 &     $1.1\mathrm{e}^{-5}$ \\
		2019 &      0.73 &     0.54 &     0.58 &     $1.5\mathrm{e}^{-17}$ \\
		2020 &      0.27 &     0.071 &     0.19 &     $5.4\mathrm{e}^{-3}$ \\
		\hline
		{} & \multicolumn{4}{c}{\textbf{Standard temporal split}}\\
		\hline
		Year &  Pearson correlation & $R^2$ &  Kendall tau & p-value of Kendall tau \\
		\hline
		2016 &      -0.43 &     0.18 &     -0.29 &     $1.6\mathrm{e}^{-5}$ \\
		2017 &      0.46 &     0.21 &     0.33 &     $9.2\mathrm{e}^{-7}$ \\
		2018 &      0.46 &     0.21 &     0.34 &     $7.7\mathrm{e}^{-7}$ \\
		2019 &      0.47 &     0.22 &     0.33 &     $1.3\mathrm{e}^{-6}$ \\
		2020 &      0.47 &     0.22 &     0.39 &     $7.6\mathrm{e}^{-9}$ \\
	\end{tabular}}
	\caption{Dependence measures between the cross-entropies (prediction error) in the validation and test sets, for the 100 best models of the random hyperparameters search.}
	\label{tab:validation_test_corr_table}
\end{table}

Our second and most important aim is to establish that ESG data contain additional valuable and exploitable information on price returns in comparison to a set of benchmark features. To this end, for each training period defined above, we train a model with both ESG and benchmark features and another model with benchmark features alone. We assess both the absolute performance metrics of the models and the extent of additional information provided by ESG features by calculating the difference in performance metrics in the test sets.

The company-wise splits make it easy to compute error bars on various metrics: instead of training $K=5$ models, we train a hundred of them and then compute the median performance on a hundred random subsets of size $K=5$ among these 100 models. Table \ref{tab:refinitivresultstable_CAPMAlpha} provides results on the absolute performance of the models for each test period for both the company-wise and the standard temporal splits. Both splitting methods have a clearly decreasing cross-entropy (a proxy for prediction error) as a function of time, except for 2020, which shows once again the special nature of this year in our dataset. This shows that the relevance of ESG features in price return formation increases as a function of time.
Balanced accuracy displays a similar improvement before 2020. However, this time, yields of company-wise splits are increasingly better than temporal splits, which we believe is an encouraging sign of its ability to better leverage the latest and best data.

\begin{figure}
	\begin{subfigure}{0.5\linewidth}
		\centering
		\includegraphics[width=0.98\linewidth]{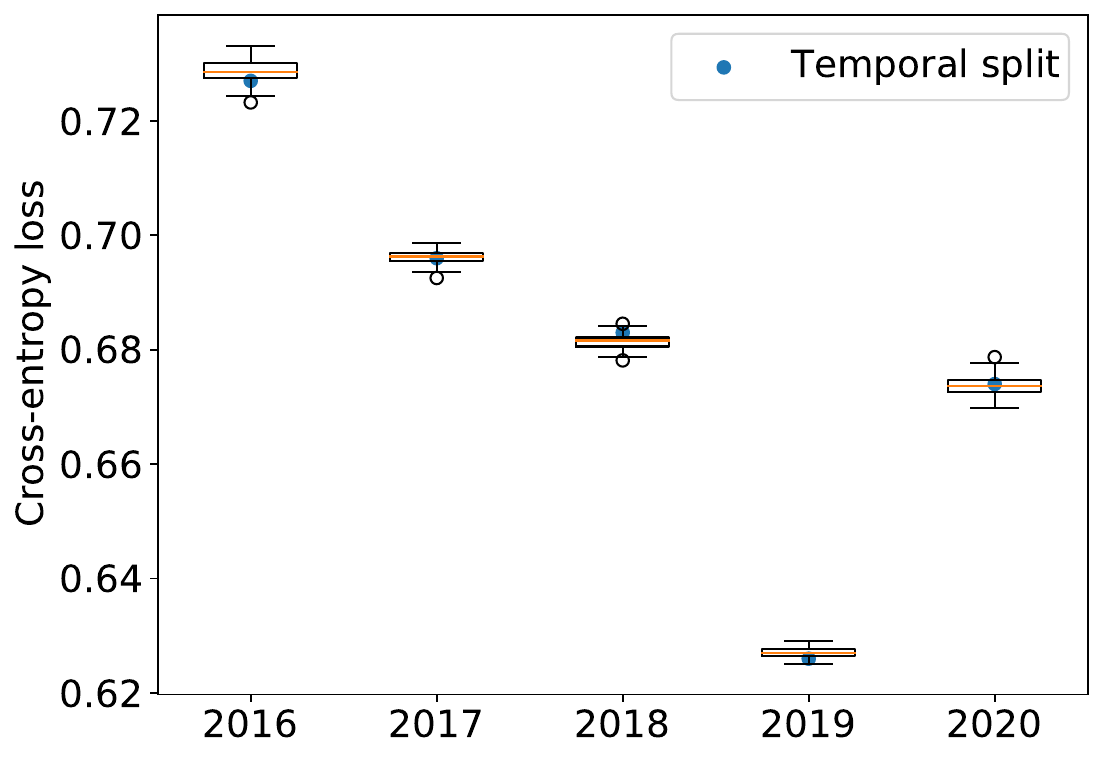} 
		\caption{Cross-entropy}
		\label{fig:cross-entropy_comparisons}
	\end{subfigure}
	\begin{subfigure}{0.5\linewidth}
		\centering
		\includegraphics[width=1\linewidth]{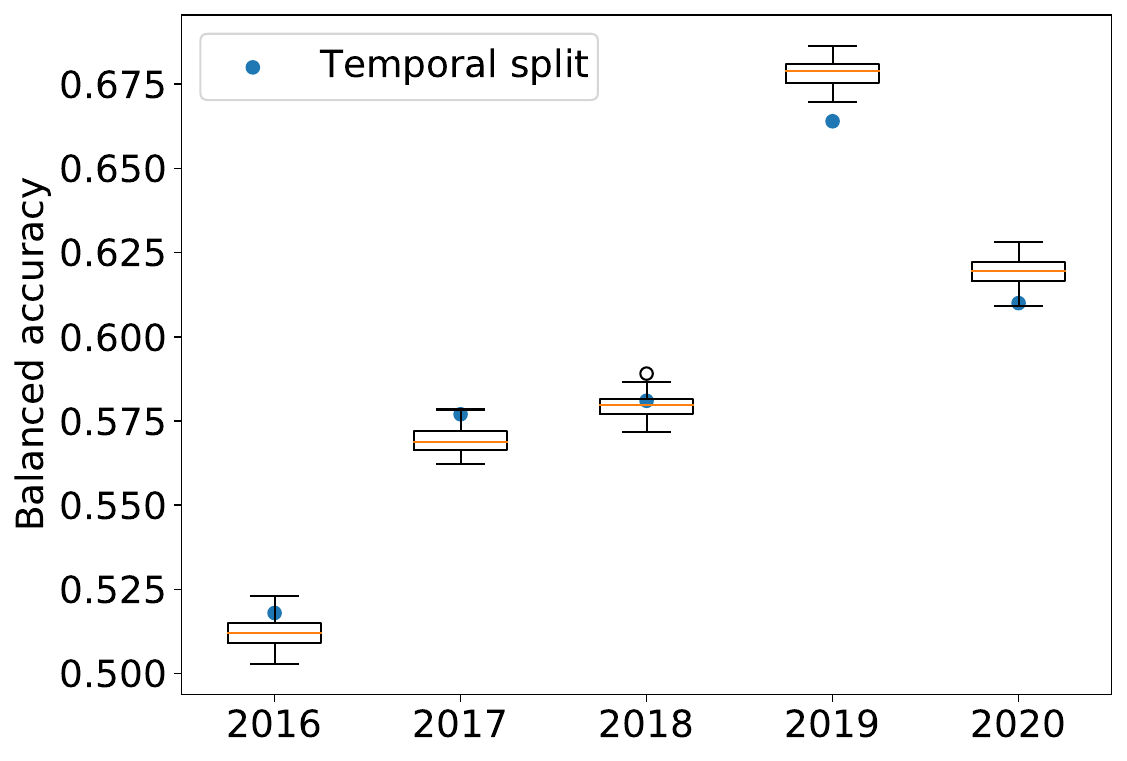}  
		\caption{Balanced accuracy}
		\label{fig:balanced_acc_comparisons}
	\end{subfigure}
	\caption{Performance measures on the test sets of the two train and validation schemes. The boxplots show the performance of 100 random samplings of 5 models among 100 random company-wise validation splits.}
	\label{fig:Comparison_split_boxplot}
\end{figure}

Figure \ref{fig:Comparison_split_boxplot} displays the time evolution of the cross-entropy and the balanced accuracy in the test sets. The boxplots are computed for the company-wise splits from the 100 associated predictions; the orange lines are the median of these performance measures, the rectangle delimits the first and third quartiles, and extreme limits are situated before the first quartile minus 1.5 times the interquartile range and after the third quartile plus 1.5 times the interquartile range. Any point outside of this range is considered an outlier. 

Company-wise $5$-fold cross-validation outperforms the standard time-splitting scheme, which supports our claim that the not fully mature nature of ESG data can be partly alleviated by a suitable validation scheme. 

\begin{table}
	\centering
    \resizebox{\textwidth}{!}{\begin{tabular}{l||c|c||c|c}
    	    {} & \multicolumn{4}{c}{\textbf{Company-wise $5$-fold cross-validation}}\\
    	    \hline
    		{} & \multicolumn{2}{c||}{Only Benchmark features} & \multicolumn{2}{c}{Benchmark and ESG features} \\
    		\hline
    		Year &   Balanced Accuracy & Cross-entropy loss &  Balanced Accuracy & Cross-entropy loss \\
    		\hline
    		2016 &      52.6 &     70.6 &     51.2 &     72.8 \\
    		2017 &      57.4 &     69.2 &     56.9 &     69.6 \\
    		2018 &      57.5 &     68.1 &     57.9 &     68.2 \\
    		2019 &      65.6 &     63.1 &     67.9 &     62.7 \\
    		2020 &      59.6 &     69.3 &     61.9 &     67.4 \\
    		\hline
    		{} & \multicolumn{4}{c}{\textbf{Standard temporal split}}\\
    		\hline
    		{} & \multicolumn{2}{c||}{Only Benchmark features} & \multicolumn{2}{c}{Benchmark and ESG features} \\
    		\hline
    		Year &   Balanced Accuracy & Cross-entropy loss &  Balanced Accuracy & Cross-entropy loss \\
    		\hline
    		2016 &      53.2 &     68.8 &     51.8 &     70.3 \\
    		2017 &      56.1 &     68.2 &     57.7 &     68.0 \\
    		2018 &      56.2 &     67.5 &     58.1 &     67.4 \\
    		2019 &      64.3 &     64.5 &     66.4 &     63.8 \\
    		2020 &      58.5 &     70.5 &     61.0 &     69.6 \\
    \end{tabular}}
	\caption{Performance measures in percent on the test set for both types of validation splits. The numbers for the company-wise splits are the median values of the performance of 100 random samplings of 5 models among 100 random company-wise validation splits.}
	\label{tab:refinitivresultstable_CAPMAlpha}
\end{table}

\begin{figure}
        \centering
		\includegraphics[width=0.7\linewidth]{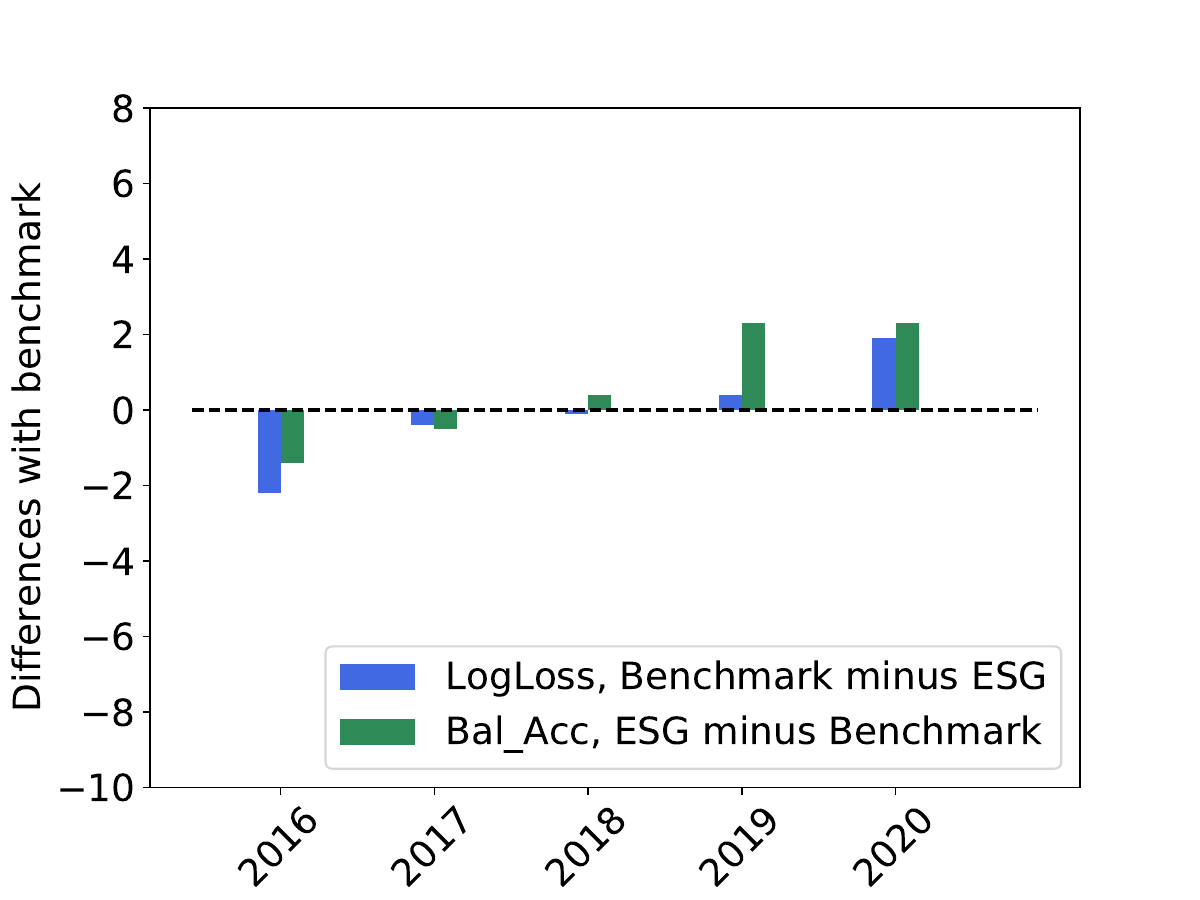}
		\caption{Performance measures in comparison to benchmark, for the company-wise $5$-fold cross-validation.}
		\label{fig:diffwithbench_CAPM_INSTS5}
\end{figure}

Figure \ref{fig:diffwithbench_CAPM_INSTS5} shows the difference in performance between the models trained on ESG and benchmark features and the models trained only on benchmark features for the company-wise $5$-fold cross-validation. ESG features contain more relevant information as time goes on. Two explanations spring to mind: long positions are more and more driven by ESG-conscious investors, or the quality of data increases as a function of time, which makes the relevance of ESG scores more apparent.

\section{Interpretability}

We now provide a breakdown of the impact of the different ESG features on the predicted probability of having positive idiosyncratic returns in the CAPM model. Because of the superior performance of the company-wise $K$-fold cross-validation, we use this method in the following.

\subsection{Shapley values}

Shapley values, first introduced in the context of game theory \citep{shapley1953value}, provide a way to characterize how each feature contributes to the formation of the final predictions. Shapley values and their uses in the context of machine learning are well described in \cite{molnar2020interpretable}.

The Shapley value of a feature can be obtained by averaging the difference in prediction between all the combinations of features containing and not containing the said feature. For each sample in our dataset, each feature possesses its own Shapley value representing the contribution of this feature to the prediction for this particular sample. Shapley values have very interesting properties, one of them being the efficiency property. If we note $\phi_{j, i}$ the Shapley value of feature $j$ for a sample $x_i$ and $\hat{f}(x_i)$ the prediction for the sample $x_i$, Shapley values must add up to the difference between the prediction for the sample $x_i$ and the average of all predictions $E_X(\hat{f}(X))$ and then follow the following formula:
\begin{align}
	\sum\nolimits_{j=1}^p\phi_j=\hat{f}(x)-E_X(\hat{f}(X))
\end{align}

The dummy property also states that the Shapley value of a feature that does not change the prediction, whatever combinations of features it is added to, should have a Shapley value of 0. 

Shapley values computation is quite time-and memory-intensive. \cite{lundberg2017unified} and later \cite{lundberg2018consistent} proposed a fast implementation of an algorithm called TreeSHAP, which allows to approximate Shapley values for trees models such as the LightGBM, which we use in the following and refer to as SHAP values.

Let us just note that, as we are using a LightGBM model in classification, the prediction is not directly the probability of belonging to class 1, but rather the logit associated with this probability. Probability is an increasing function of the logit, and thus, SHAP values obtained for the logit can easily be transformed for the probability. Indeed, for a sample $x_i$, the predicted probability of belonging to class 1 $p_i$ is linked to the logit $\textrm{logit}_i$ according to:\vspace{6pt}
\begin{align}
	p_i = \frac{1}{1 + e^{-\textrm{logit}_i}}
\end{align}

\subsubsection{Evolution of ESG features contribution from 2017 to 2020}
\label{sub:evolutionshap}

\begin{figure}
	\begin{subfigure}{0.5\linewidth}
		\centering
		\includegraphics[width=1\linewidth]{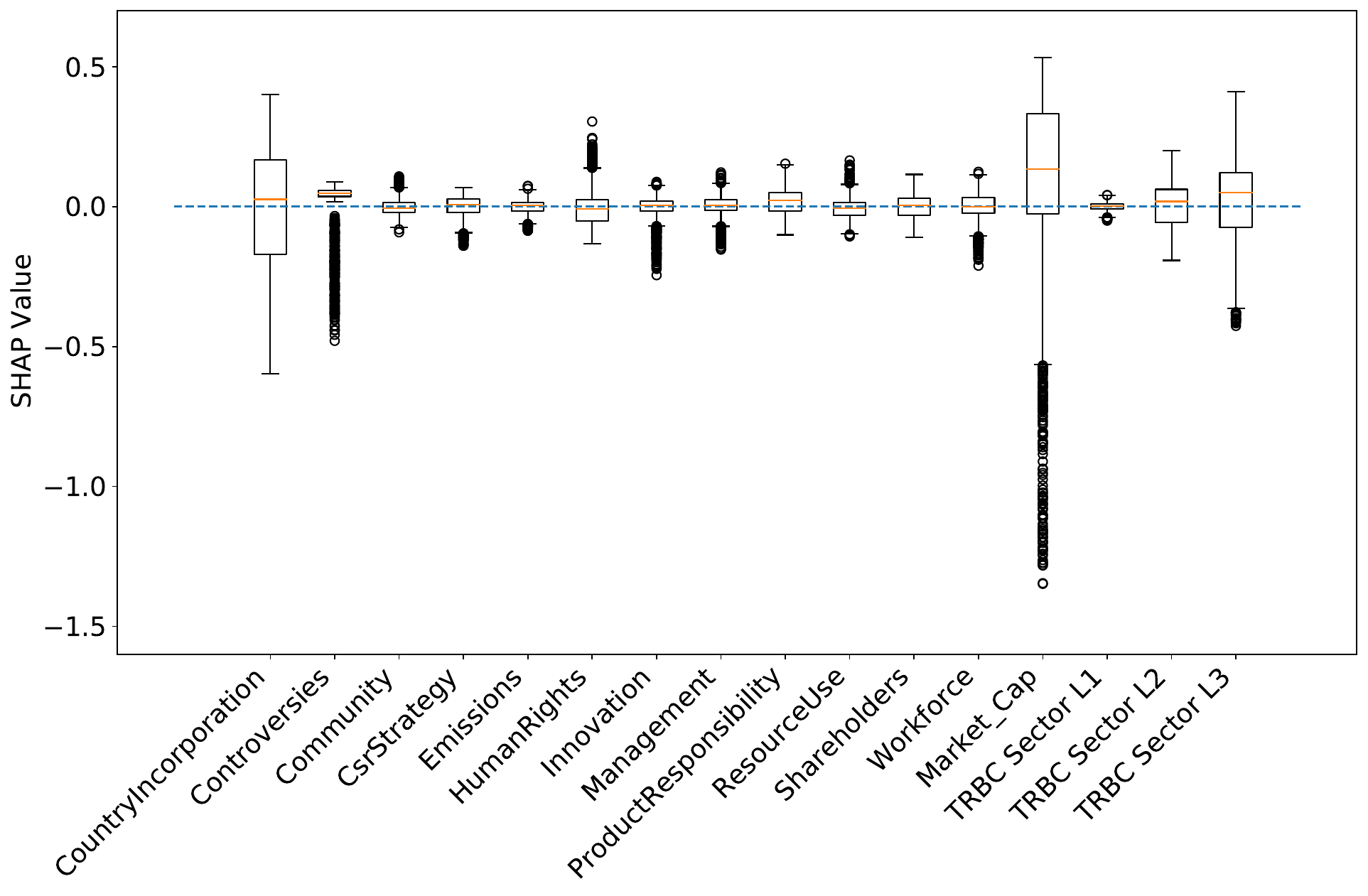}  
		\caption{Test year 2017}
		\label{fig:shap2017}
	\end{subfigure}
	\begin{subfigure}{0.5\linewidth}
		\centering
		\includegraphics[width=1\linewidth]{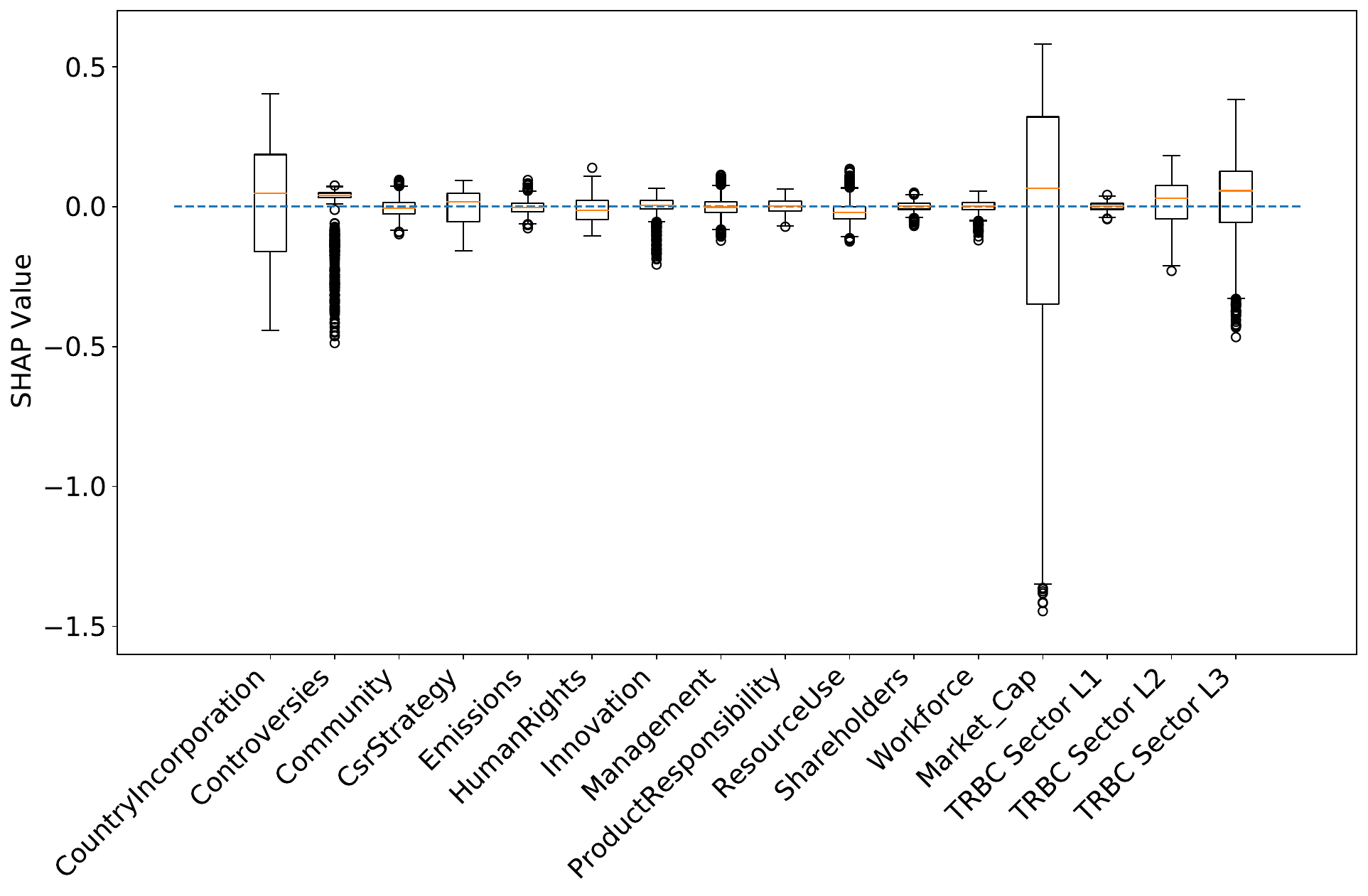}  
		\caption{Test year 2020}
		\label{fig:shap2020}
	\end{subfigure}
	\caption{SHAP values distribution.}
	\label{fig:shap}
\end{figure}

In Figure \ref{fig:shap}, we plot the distribution of SHAP values for each feature and for all test samples for models trained from 2002 to 2016 (Figure \ref{fig:shap2017}) and trained from 2002 to 2019 (Figure \ref{fig:shap2020}). The first teaching of this plot is that the contribution of ESG features to the predicted probability of having a positive return has not dramatically increased with the additional, more recent and more complete data. Benchmark features are the ones that have the biggest impact on the prediction. However, we observe an important number of outliers for some SHAP values associated with some features, demonstrating that these ESG features have more impact on the prediction for these particular samples. It would be interesting to study these outliers to understand more why ESG features are more important in explaining price returns for some samples than others.

\begin{figure}
	\centering
	\includegraphics[width=0.5\linewidth]{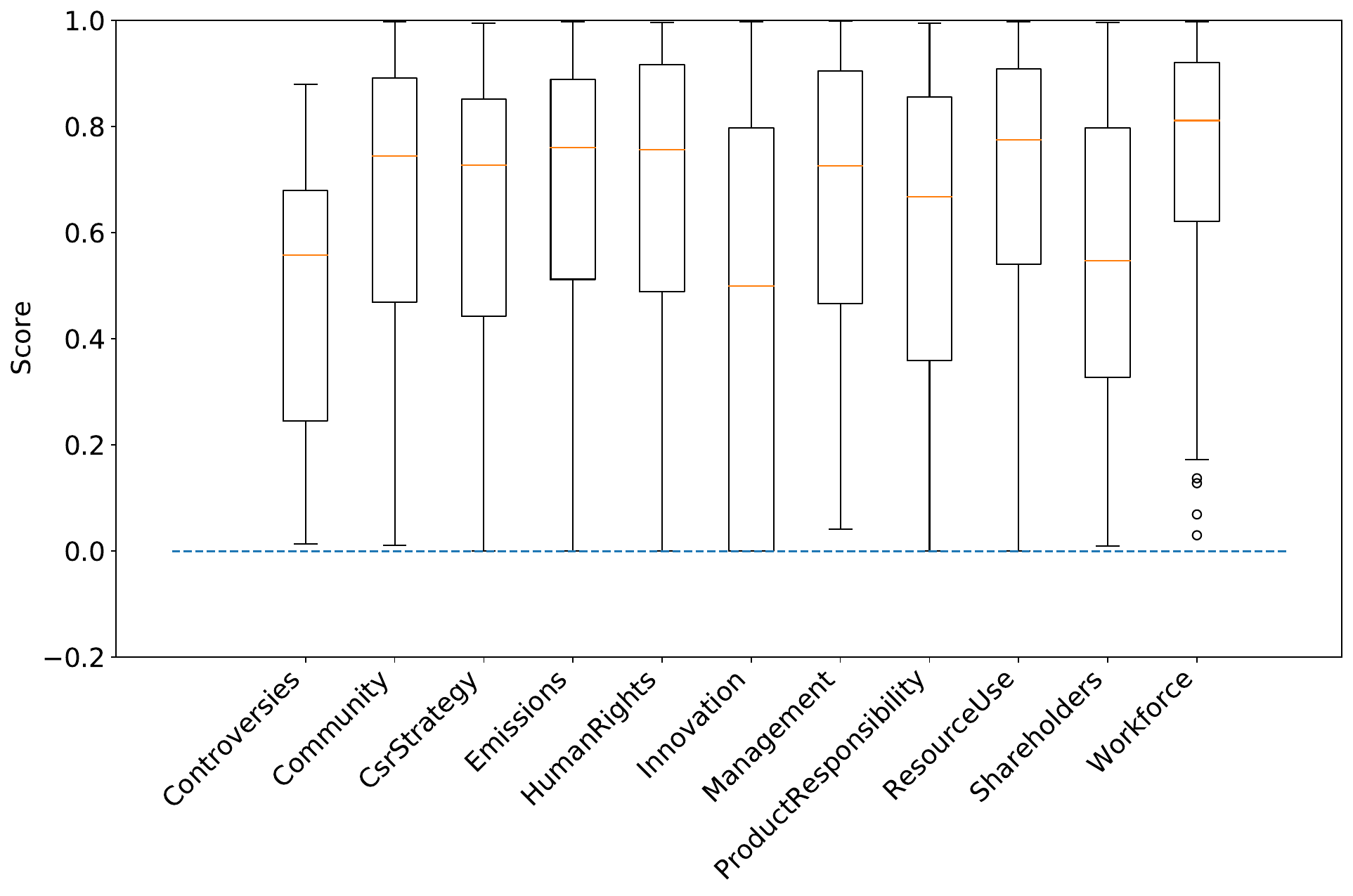}  
	\caption{Distribution of data for lowest outliers of SHAP values of 2020, for Controversy score.}
	\label{fig:datadistribcontrovoutliersshap2020}
\end{figure}

For instance, we observe in Figure \ref{fig:datadistribcontrovoutliersshap2020} the score distributions for the outliers of the Controversy SHAP values. All of these scores are below 0.9, suggesting that the Controversy score is more informative when a company has indeed suffered controversies during the year and was then not able to reach a score of 1. Observing outliers of SHAP values and their associated scores, we can make the hypothesis that ESG features are important and have a strong impact on the explanations of past returns if their score is extreme. This would mean that ESG information would lie in extreme scores, with more standard scores bringing much less information. Checking this hypothesis is beyond the scope of this work and is left for future investigations.

\subsubsection{On the choice of SHAP values as the interpretation method}

The use of SHAP values enables the computation of feature-specific explanations for individual samples, providing insight into the contribution of each feature to a given prediction. This approach is founded on sound theoretical principles and constitutes an exact method, as indicated by \cite{molnar2020interpretable}: the cumulative sum of all SHAP values yields the predicted logit, and subsequently, the predicted probability outputted by the model can be retrieved.  TreeSHAP implementation is a fast method for computing SHAP values. Although other techniques for explanatory purposes exist, they lack the beneficial properties inherent to SHAP values. We proceed to discuss two such methods, namely the feature importance derived from the LightGBM model and Local Interpretable Model-Agnostic Explanations (LIME).

As a gradient-boosted model, LightGBM allows the derivation of feature importance metrics directly from the trained model. Typically, this is computed as the number of times a particular feature is utilized in building each of the trees comprising the LightGBM model. However, such an approach to computing feature importance yields only a global measure and does not allow the derivation of per-sample feature importance. This shortcoming renders it impossible to ascertain if a given feature can yield divergent effects depending on the sample being evaluated, as shown by the SHAP values presented in Figure \ref{fig:shap}. Consequently, the feature importance derived from LightGBM is usually deemed unsatisfactory for explaining models due to its lack of granularity.

Other machine learning interpretation methods that provide a per-sample explanation are available. One such method is LIME, developed by \cite{ribeiro2016should}. \cite{molnar2020interpretable} discusses this methodology, highlighting its advantages and drawbacks. Although LIME provides per-sample explanations, it does not constitute an exact method and lacks robustness. Additionally, the explanations derived via LIME are unstable, as they can vary upon running the methodology on different occasions. Moreover, when the model being explained is not locally linear, the LIME methodology is not meaningful. These drawbacks are illustrated in the work of  \cite{alvarez2018robustness}, which discusses the limitations of LIME with respect to robustness.

\subsection{Partial dependence plots: marginal effect of ESG features}

\subsubsection{Definition}

A partial dependence plot shows the marginal effect of features on the prediction made by the model. It is a way of understanding the links the model made from features to the target and that it had understood from the data. It also shows if this relation is linear or not, monotonic or not, etc. Partial dependence plots were first introduced by \cite{friedman2001greedy} and are also well-described by \cite{molnar2020interpretable}.  Briefly, a partial dependence plot for a feature of interest is obtained by marginalizing the predicted output over the values of all other input features. This marginalization is performed by calculating averages in the training data, using a Monte-Carlo method, with a fixed value for the features of interest.

An important limitation of a partial dependence plot is that their methodology of construction assumes independence between the features, which does not seem to be the case for ESG features. This limitation is neglected here. All partial dependence plots are made with the most recent model, trained with data from 2002 to 2019, on a subsample of recent ESG data.

\subsubsection{Marginal effect of ESG features}

\begin{figure}
	\centering
	\includegraphics[width=0.95\linewidth]{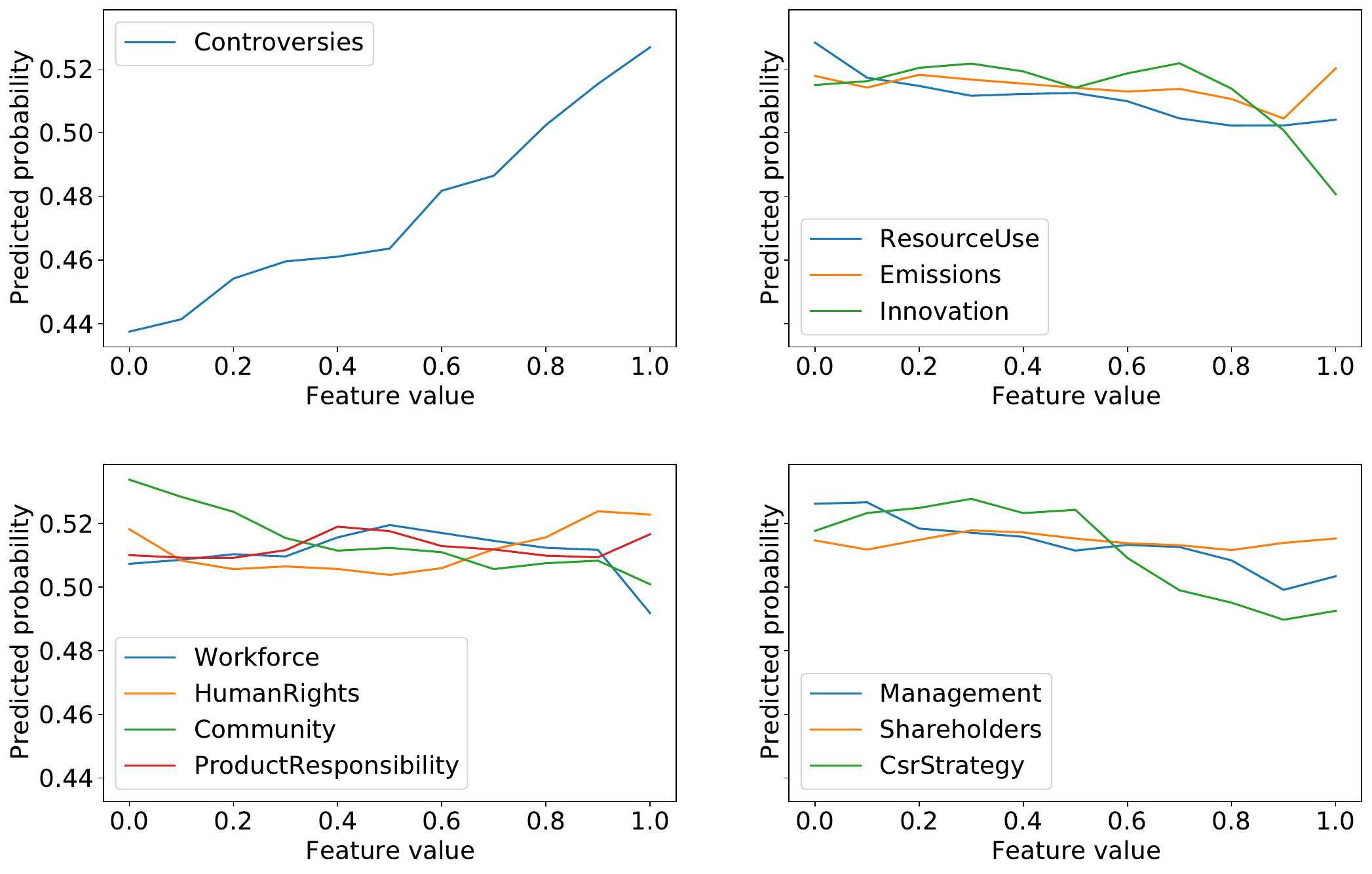}  
	\caption{Marginal effect of each ESG feature on the predicted probability of having a positive return.}
	\label{fig:PDP_F1}
\end{figure}

\begin{figure}
	\centering
	\includegraphics[width=0.85\linewidth]{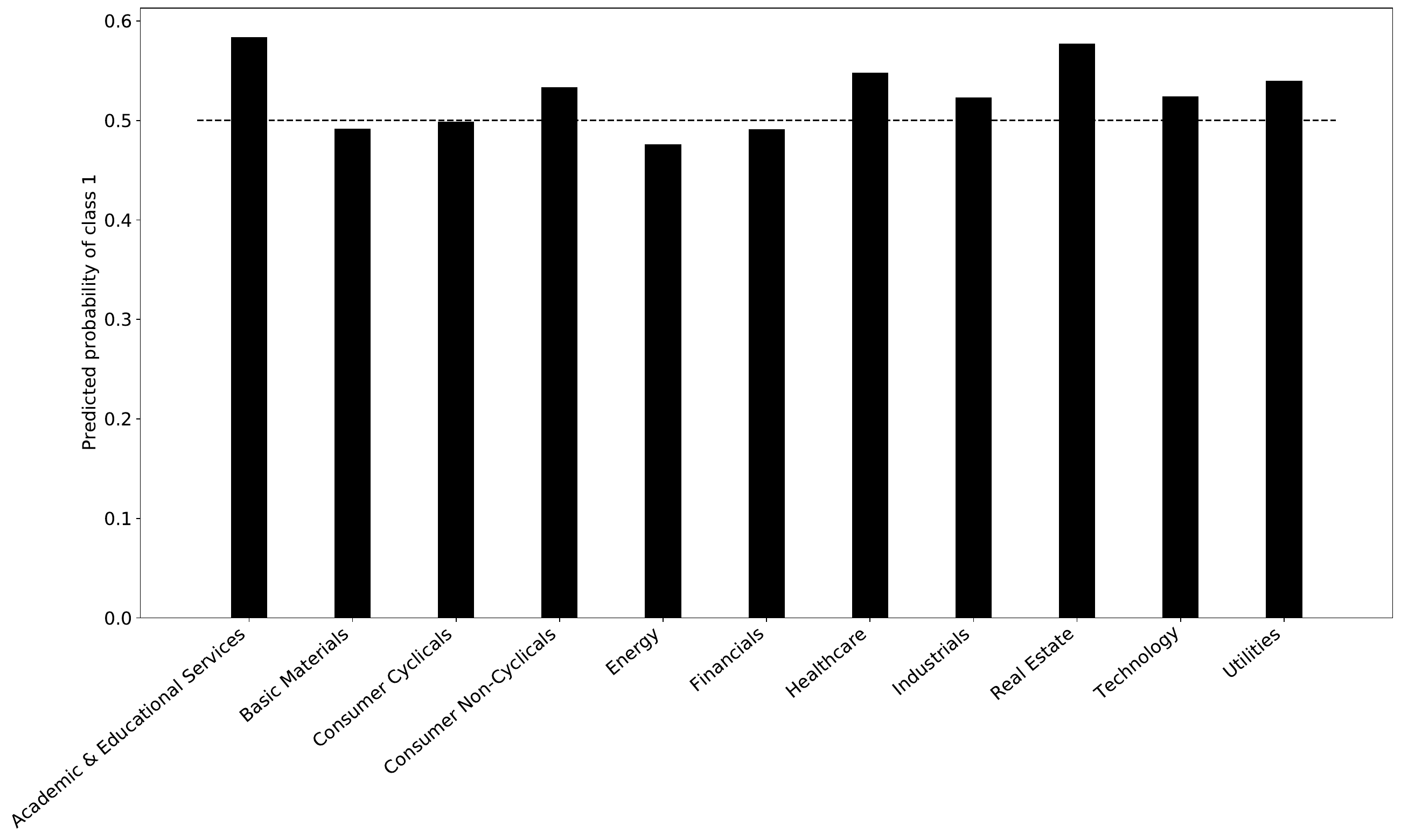}  
	\caption{Marginal effect of the sector (TRBC Sector L1) feature on the predicted probability of having a positive return.}
	\label{fig:PDP_F1_TRBCL1}
\end{figure}

Using partial dependence plots, we first compute the marginal effect of each ESG feature on the probability of having a positive return during the year of publication of the ESG features (Figure \ref{fig:PDP_F1}). Figure \ref{fig:PDP_F1_TRBCL1} reports the sector by sector probability of having a positive predicted return.

Figure \ref{fig:PDP_F1} shows that ESG features are mostly not related in a monotonic way with the probability of having a positive return. A clear exception would be the Controversy score, on the top left, which shows a strong monotonic relation and strongly implies that  being subject to controversies during a year leads to a lower probability of having a positive return. For the 10 pillar scores, one sees a much weaker dependence. For example, the probability of positive price return increases by around 1\% when the Product Responsibility and Shareholders scores increase from 0 to 1. Still, a trend is present for most of these ESG features: partial dependence plots for features such as  Resource Use, Innovation, Community or Management seem to be decreasing, suggesting that obtaining better ESG scores and practices comes at the price of a slightly degraded financial performance.

\subsubsection{Marginal effect of ESG Features sector by sector: materiality matrices}

Adding the sector dimension to partial dependence plots yields so-called materiality matrices. In our setting, it is a table whose rows represent ESG features and whose columns are economic sectors. A cell of this matrix shows, in percentage, by how much the probability of having a positive return is increased by going from a low score (between 0 and 0.2) to a high one (by 0.8 to 1). This quantity is easily obtained using partial dependence plots: for a specific selected economic sector, we can plot the evolution of the predicted probability against the feature value. Making the strong hypothesis of a monotonic and close-to-linear relationship, we can compute the value in the cell as the slope of the trend line of the precedent plot.

\begin{figure}
	\centering
	\includegraphics[width=0.8\linewidth]{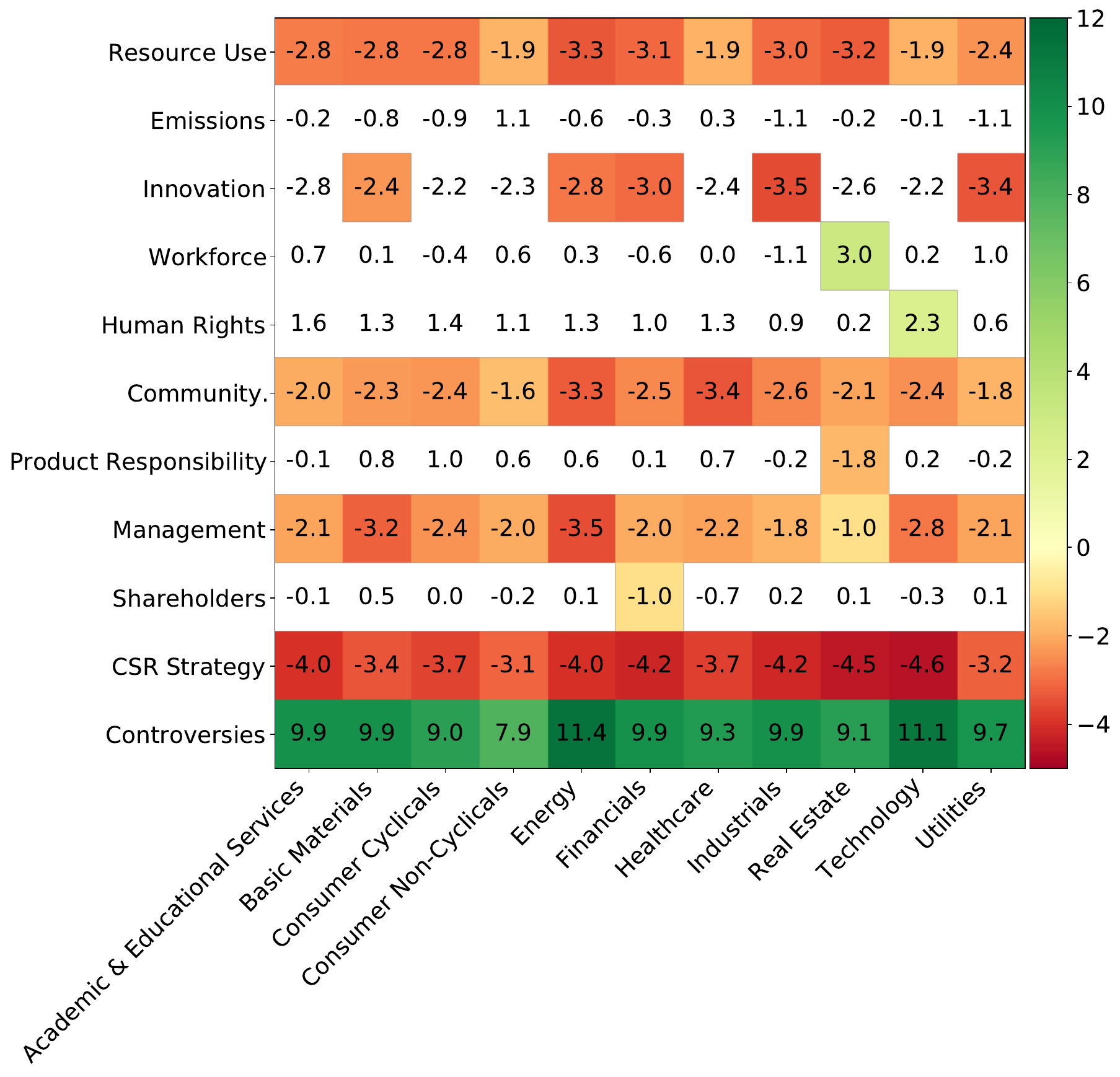}  
	\caption{Materiality matrix: marginal effects of the combination ESG feature/Sector feature on the predicted probability of having a positive return. Blank cells are those which were not found statistically significant by the Benjamini-–Hochberg procedure.}
	\label{fig:PDP_F2_allcap}
\end{figure}

The obtained materiality matrix is presented in Figure \ref{fig:PDP_F2_allcap}. All the TRBC sectors of level 1 are included. Results for Academic and Educational Services should be handled with care as they are not based on as many samples as the ones for other sectors, as shown in Figure \ref{fig:sampleseurperl1eur}. Some ESG scores have a  strong impact on the probability of having positive returns. The Controversy score especially has a similar impact for all sectors: not suffering controversies during the year increases the probability of having a positive return. On the contrary, the CSR Strategy row shows that working towards the integration of social and environmental dimensions into the day-to-day decision-making processes, in addition to economic and financial ones, leads to a loss of financial performance. It is also the case for  Resource Use, Environmental Innovation, Community, and Management scores, each with a different magnitude.

\begin{figure}
	\begin{subfigure}{0.5\linewidth}
		\centering
		\includegraphics[width=1\linewidth]{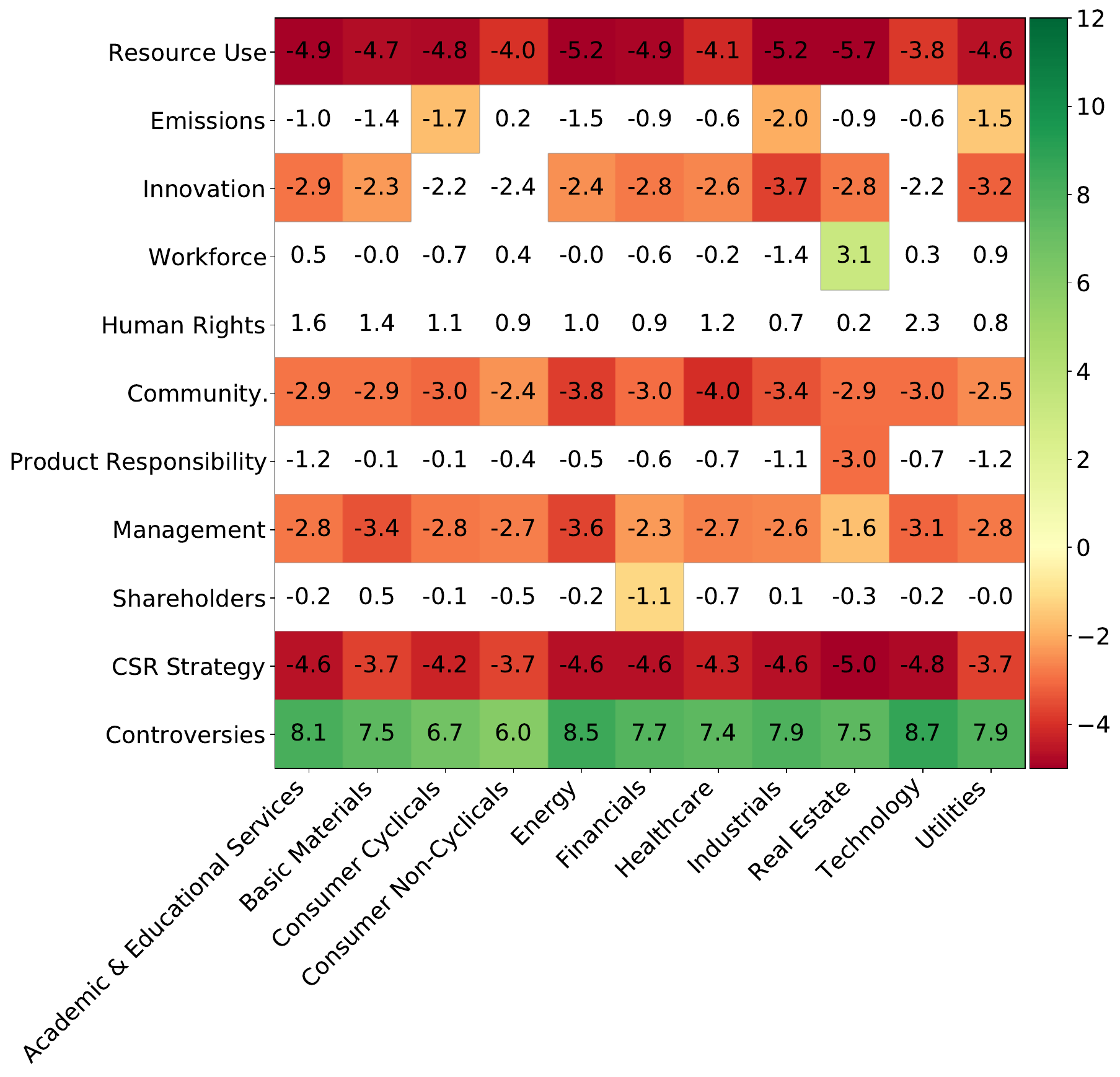}  
		\caption{Small market capitalization (<2B€)}
		\label{fig:PDP_F2_smallcap}
	\end{subfigure}
	\begin{subfigure}{0.5\linewidth}
		\centering
		\includegraphics[width=1\linewidth]{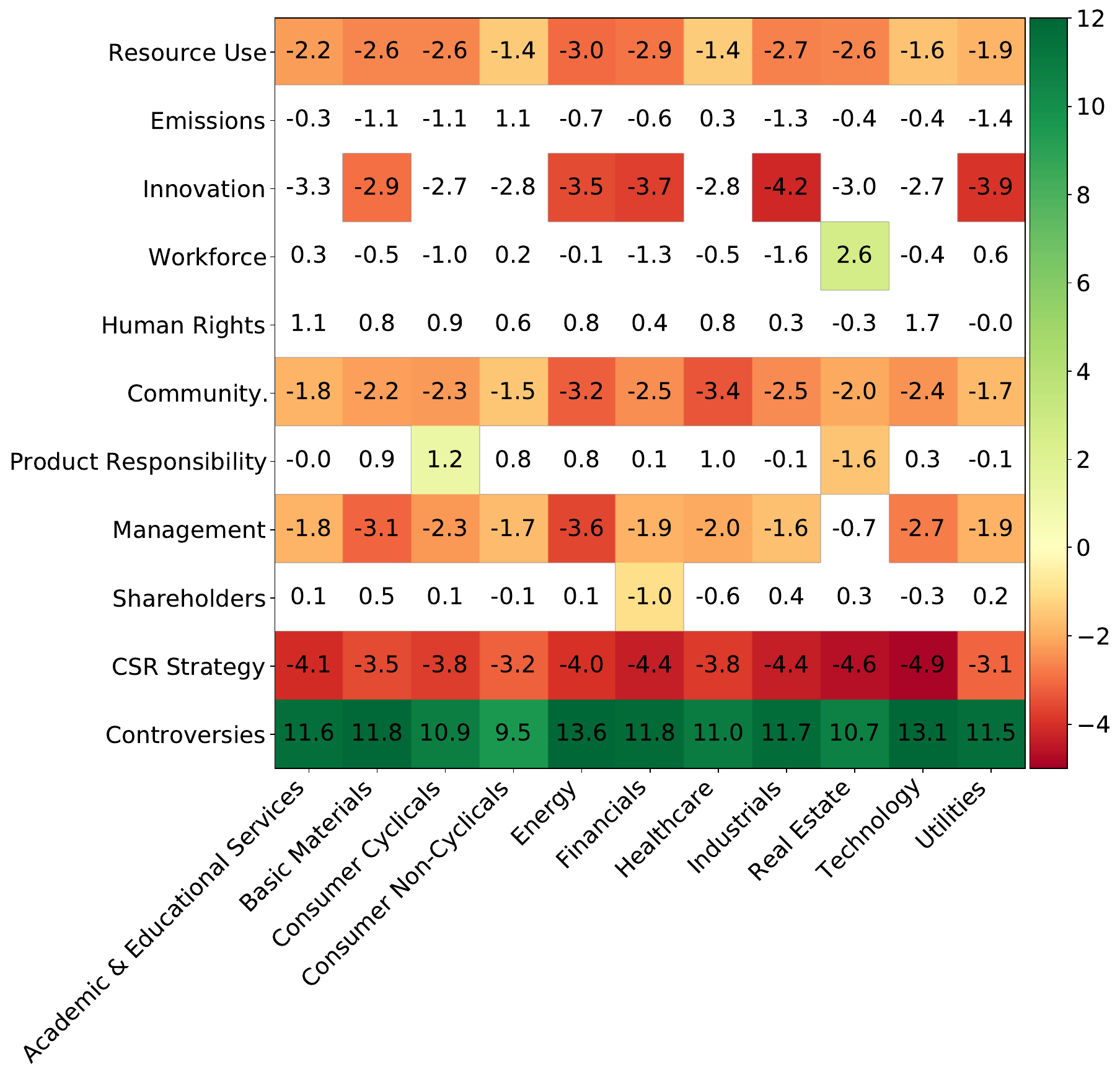}  
		\caption{Mid market capitalization (>2B€, <10B€)}
		\label{fig:PDP_F2_midcap}
	\end{subfigure}
	\begin{subfigure}{\textwidth}
		\centering
		\includegraphics[width=0.5\linewidth]{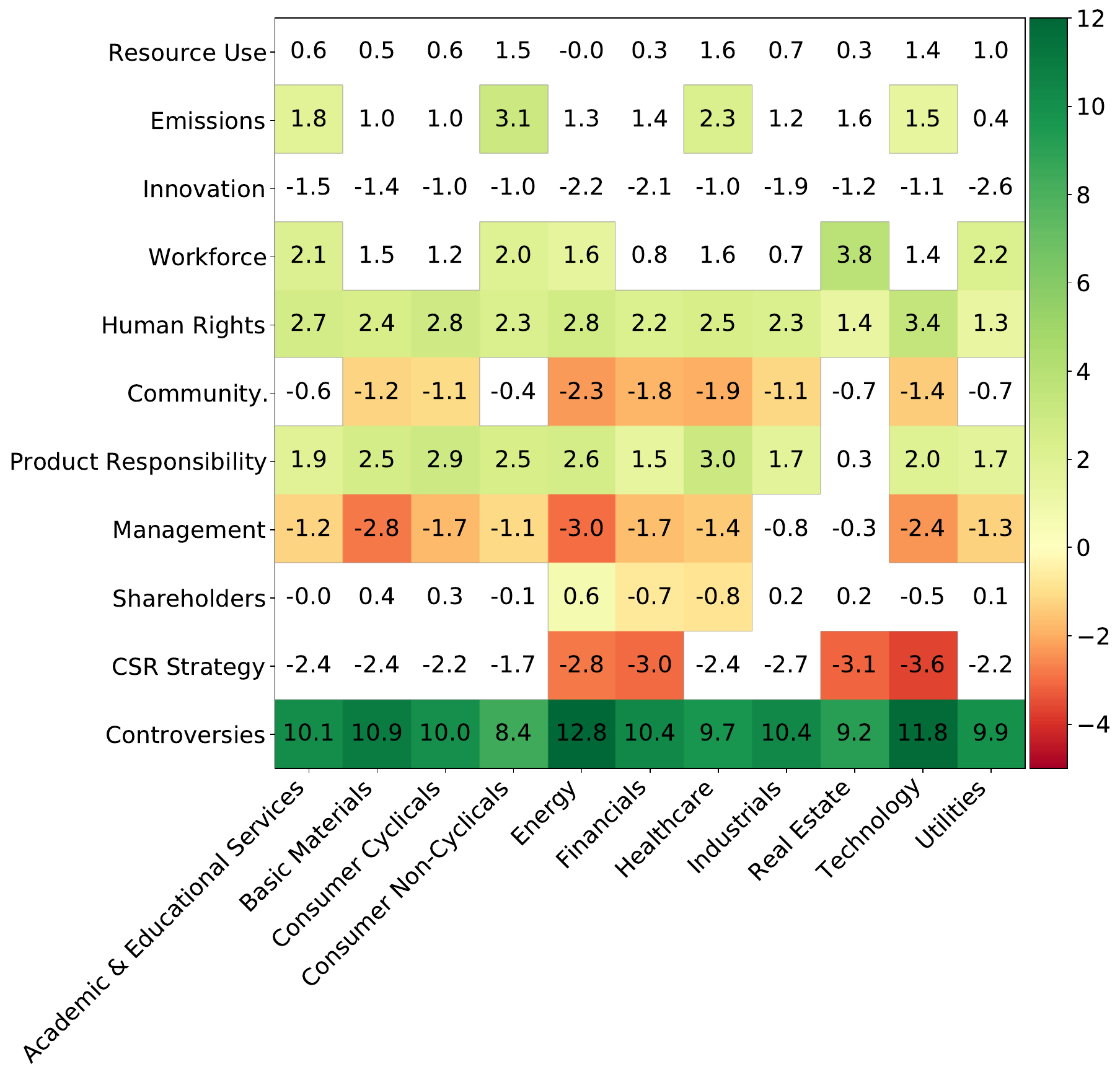}  
		\caption{Large market capitalization (>10B€)}
		\label{fig:PDP_F2_largecap}
	\end{subfigure}
	\caption{Materiality matrices: marginal effects of the combination ESG feature/Sector feature on the predicted probability of having a positive return, bucketed by market capitalization. Blank cells are those which were not found statistically significant by the Benjamini–-Hochberg procedure.}
	\label{fig:PDP_F2_bycap}
\end{figure}

Furthermore, we bucket the companies that serve to build this materiality matrix by market capitalization. We choose three buckets, with small market capitalization being below 2 billion euros, mid ones between 2 and 10 billion euros and large ones above \mbox{10 billion euros}, which correspond to the ones Refinitiv uses when calculating the Controversy score. The three obtained materiality matrices are presented in Figure \ref{fig:PDP_F2_bycap}. The marginal effect of the Controversy score remains the same, even if it is slightly smaller for the small caps. However, companies with a large market capitalization benefit from a better impact of ESG: for some features, working toward better ESG scores can preserve or even boost financial performance, whereas it would be the opposite for small caps. For instance, large cap companies have an average materiality of 0.8 for the Resource Use score and 1.5 for the Emissions score, whereas small caps ones have respectively average scores of $-$4.6 and $-$1.1, denoting a clear difference.

To obtain a statistically meaningful interpretation of these results, we need to account for the fact that  each cell corresponds to coefficients of a linear fit with associated p-values, i.e., one makes one null hypothesis per cell. We thus need to use multiple hypothesis correction to check globally which cells show statistically significant results. Here, we choose to control the False Discovery Rate with the Benjamini--Hochberg procedure \citep{benjamini1995controlling}. We set the FDR to 5\%, which means that there are only about three false discoveries in each of the reported tables.

\section{Conclusions}

While ESG data are not yet fully mature and lack long enough quality records to be amenable to easy conclusions, powerful machine learning and validation techniques make it already possible to show that they do influence yearly price returns, and increasingly so: ESG features successfully explain the part of annual price returns not accounted for by the market factor. By breaking down their influence sector-by-sector, subscore-wise and according to market capitalization, we have demonstrated that an average approach will fail to be informative. Our findings indicate that the relationship between controversies and price return is the most robust one. The average influence of all the other ESG scores significantly depends  on the market capitalization of a company: strikingly, most of the statistically significantly influential ESG scores weigh negatively on the price returns of small or mid-size companies. Large-capitalization companies, on the other hand, have significantly advantageous ESG score types.

Our findings are specific to the Refinitiv ESG dataset for the European market, and caution should be exercised in generalizing them to other ESG datasets. This is due to the possibility of disparate ESG scores resulting from different methodological approaches to construction and from the inclusion of varying types of information. Furthermore, our study demonstrates the capacity of ESG features to provide supplementary information to explain the fraction of annual price returns not accounted for by the market factor, compared to a predetermined set of benchmark features.  The benchmark features selection was tailored to the purpose of this study: alternative choices of benchmark features could have uncovered other types of additional information embodied  in ESG features. 

While this work focuses on explaining the sign of excess price returns derived from the CAPM model, those derived from the Fama--French 3-factor model lead to results that are less clear-cut for the time being. However, this effect seems to be weakening over time: correlations between validation and test set errors increased in both 2018 and 2019, indicating the increasing information value of ESG data in explaining price returns. Future investigations will focus on the study of the full 2020 and 2021 years to verify these initial findings. Moreover, extending this research to the study of the explanatory power of ESG data with respect to more equity factors, such as quality, would enhance its comprehensiveness.

In this work, we applied  a methodology to explain the sign of price returns contemporary with ESG features.  Future research could focus on using the same framework to evaluate the predictive power of ESG data by estimating the sign of future excess returns. To achieve this objective, a distinct dataset would be necessary, containing so-called ``point-in-time'' ESG features, wherein data are not adjusted after their publication.

Future work will also include studying outliers of the SHAP values distribution and testing the hypothesis that extreme scores in the ESG field are more informative. In addition, the link between ESG and equity returns is complete only if the systematic and idiosyncratic aspects of risks and returns are studied together \citep{giese2019weighing}: indeed, it may be that having better ESG scores not only decreases price returns but also reduces risk. Future research will concentrate on investigating the information content of ESG datasets to evaluate risk measures concerning a company's stock, such as volatility or drawdown. This would provide a more comprehensive understanding of the interplay between ESG factors, risk, and equity returns.

\pagebreak
\clearpage

\bibliographystyle{plainnat}
\bibliography{reference}

\pagebreak
\clearpage

\appendix

\section{10 Pillar Scores}
\label{app:10pillarscores}

\subsection{Environmental scores}

\begin{itemize}
	\item Resource Use:	Reduce the use of natural resources and find more eco-efficient solutions by improving supply chain management.
	\item Emissions: Commitment and effectiveness towards reducing environmental emissions in the production and operational processes.
	\item Innovation: Reduce the environmental costs for customers, thereby creating new market opportunities through new environmental technologies and processes or eco-designed products.
\end{itemize}

\subsection{Social scores}

\begin{itemize}
	\item Workforce: Job satisfaction, healthy and safe workplace, maintaining diversity and equal opportunities, development opportunities for workforce.
	\item Human Rights: Respecting the fundamental human rights conventions.
	\item Community: Commitment towards being a good citizen, protecting public health and respecting business ethics.
	\item Product Responsibility: Producing quality goods and services integrating the customer's health and safety, integrity and data privacy.
\end{itemize}

\subsection{Governance scores}

\begin{itemize}
	\item Management: Commitment and effectiveness towards following best practice corporate governance principles. Composition, remuneration, transparency of the board.
	\item Shareholders:	Equal treatment of shareholders, use of anti-takeover devices.
	\item CSR Strategy:	Integration of social and environmental dimensions into the day-to-day decision-making processes, in addition to economic and financial ones.
\end{itemize}

\section{Results with the target derived from the Fama-French 3-factor model}
\label{app:targetstests}

\begin{figure}
	\begin{subfigure}{0.33\linewidth}
		\centering
		\includegraphics[width=1\linewidth]{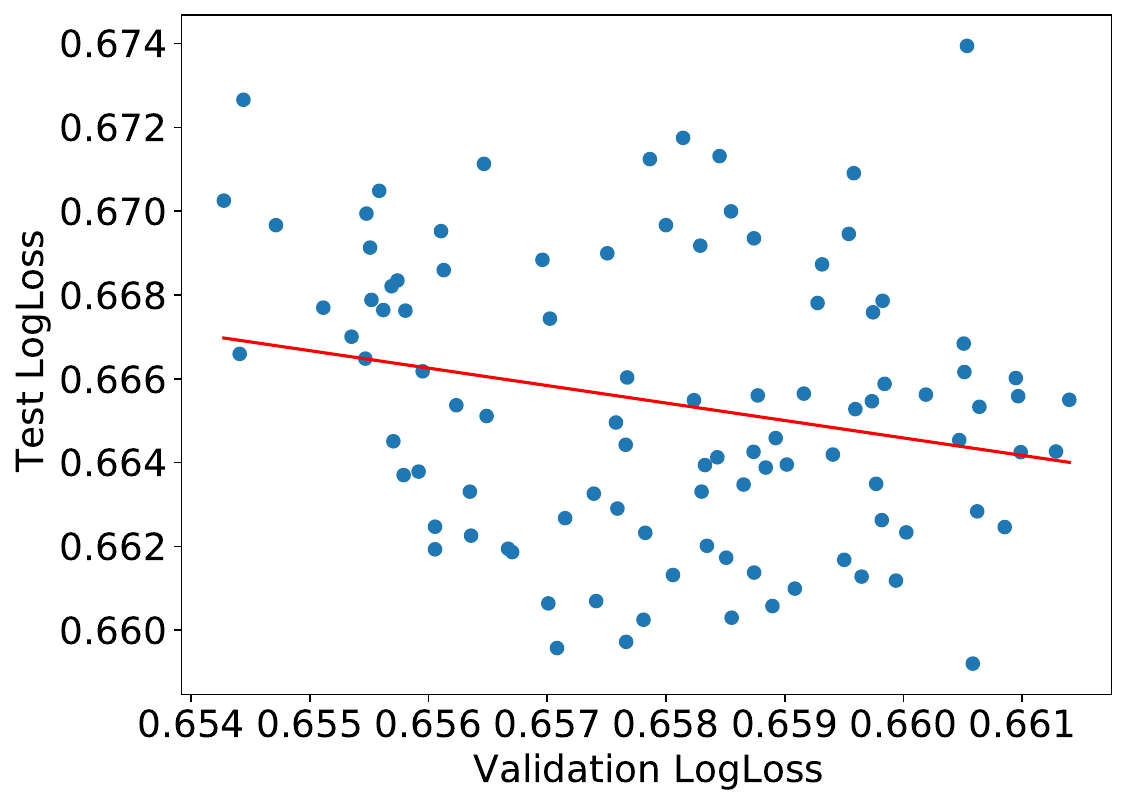} 
		\caption{2016}
		\label{fig:FF3_INST_S5_42370_-0.42_0.05}
	\end{subfigure}
	\begin{subfigure}{0.33\linewidth}
		\centering
		\includegraphics[width=1\linewidth]{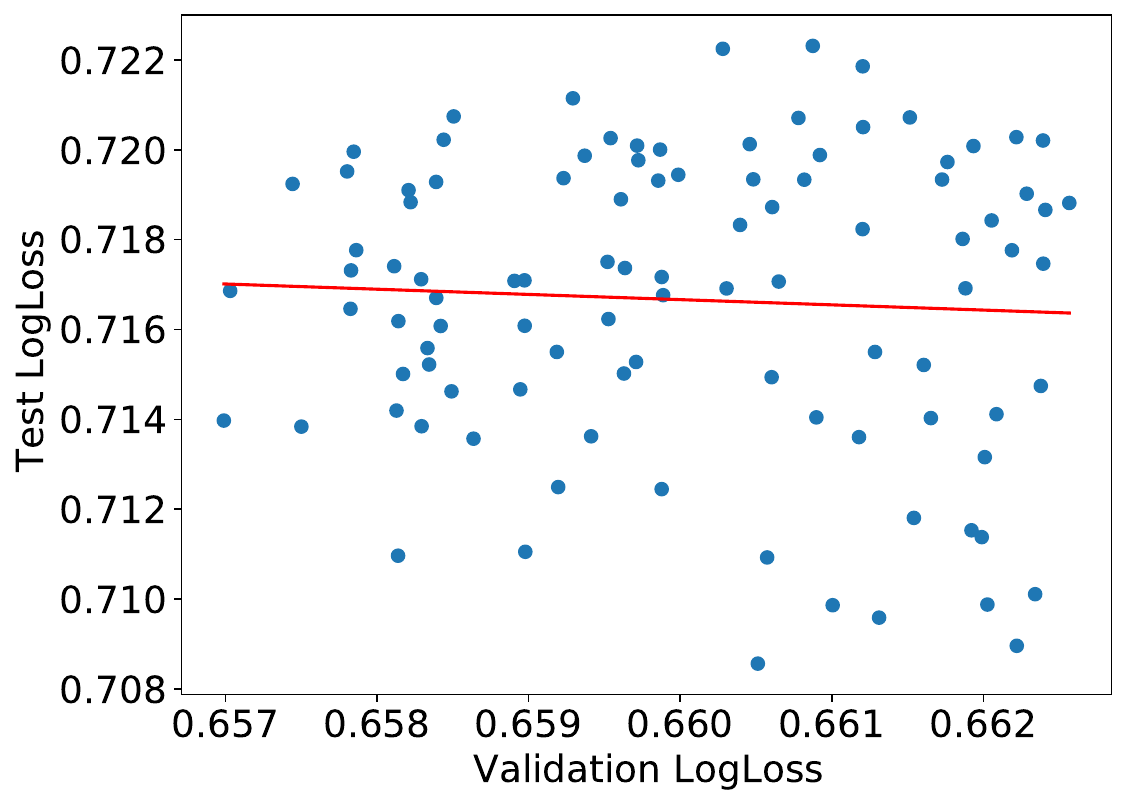}
		\caption{2017}
		\label{fig:FF3_INST_S5_42736_-0.12_0.0}
	\end{subfigure}
	\begin{subfigure}{0.33\linewidth}
		\centering
		\includegraphics[width=1\linewidth]{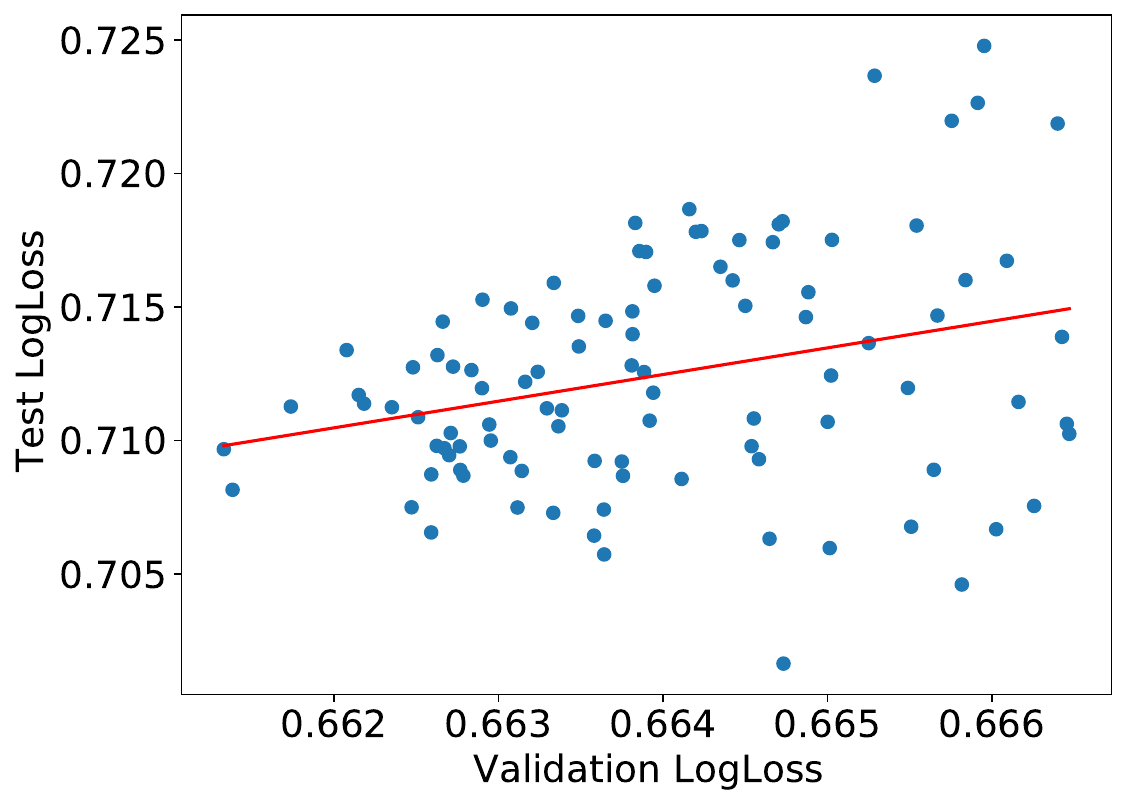}
		\caption{2018}
		\label{fig:FF3_INST_S5_43101_1.0_0.08}
	\end{subfigure}
	\begin{subfigure}{0.33\linewidth}
		\centering
		\includegraphics[width=1\linewidth]{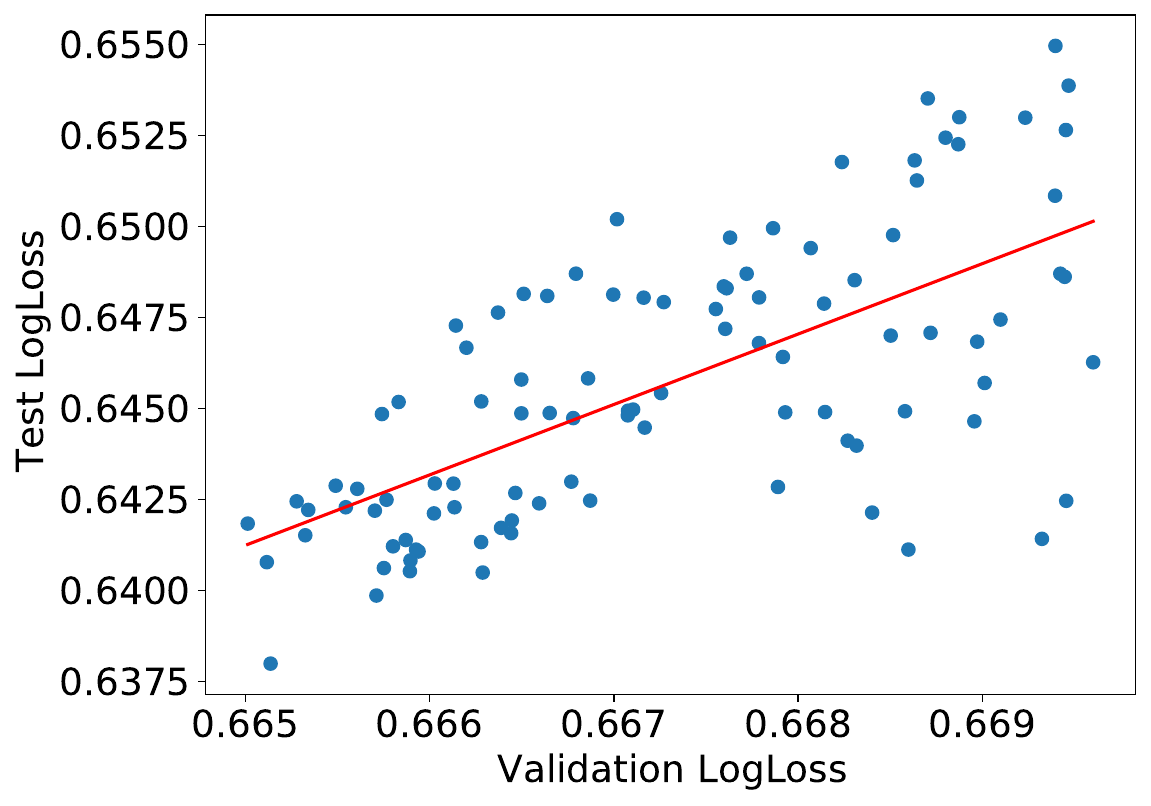}
		\caption{2019}
		\label{fig:FF3_INST_S5_43466_1.93_0.44}
	\end{subfigure}
	\begin{subfigure}{0.33\linewidth}
		\centering
		\includegraphics[width=1\linewidth]{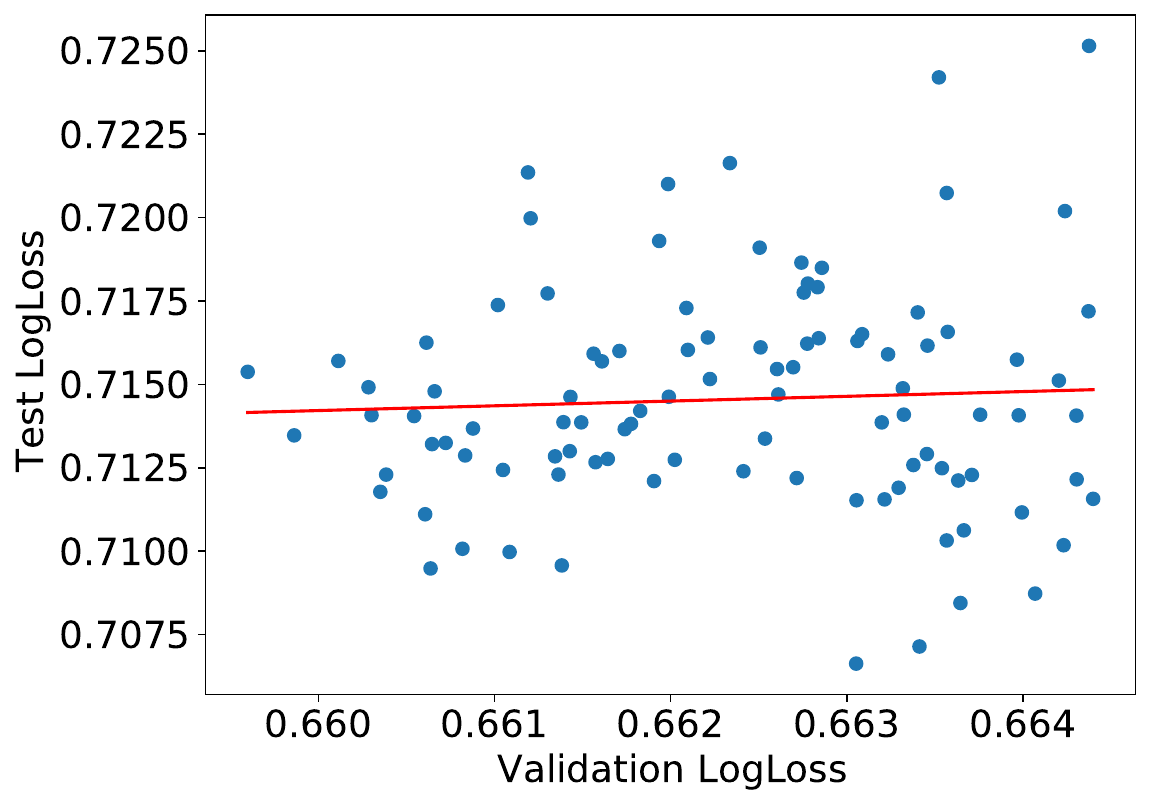} 
		\caption{2020}
		\label{fig:FF3_INST_S5_43831_0.14_0.0}
	\end{subfigure}
	\caption{Company-wise cross-validation: test set cross-entropy versus validation cross-entropy of the 100 best models of the random hyperparameters search, for a target computed using the Fama-French 3-factor model.}
	\label{fig:refinitivresultsgraph_FF3_INST_S5}
\end{figure}

The following results were obtained with a target derived from the Fama--French 3-factor model, as exposed in Section \ref{sub:problemsettings}. This target was not selected as the results were not as good as those obtained with the target derived from the CAPM model. How to interpret results with this target, especially in terms of materiality matrices, was also less clear. 
For the interested reader, we present our results, using a $5$-fold company-wise splitting strategy, in Tables \ref{tab:refinitivresultstable_FF3_INST_S5_corr} and \ref{tab:refinitivresultstable_FF3_INST_S5}. Displays of the relationship between $\mathcal{L}_m^\textrm{test}$ versus $\mathcal{L}_m^\textrm{validation}$ for each model $m$ ranking in the top 100 validation cross-entropy losses are shown in Figure \ref{fig:refinitivresultsgraph_FF3_INST_S5}.

\begin{table}
	\centering
	\resizebox{\textwidth}{!}{\begin{tabular}{l||c|c|c|c}
		{} & \multicolumn{4}{c}{\textbf{Company-wise $5$-fold cross-validation}}\\
		\hline
		Year &  Pearson correlation & $R^2$ &  Kendall tau & p-value of Kendall tau \\
		\hline
		2016 &      -0.23 &    0.052 &     -0.14 &     $4.4 \time 10^{-2}$ \\
		2017 &      -0.054 &     0.0030 &     0.010 &     $8.8 \time 10^{-1}$ \\
		2018 &      0.29 &     0.085 &     0.19 &     $4.2 \time 10^{-3}$ \\
		2019 &      0.67 &     0.44 &     0.49 &     $7.1 \time 10^{-13}$ \\
		2020 &      0.053 &     0.0028 &     0.017 &     $8.0 \time 10^{-1}$ \\
	\end{tabular}}
	\caption{Dependence measures between the cross-entropy losses in the validation and test sets, for the 100 best models of the random hyperparameters search, for a target computed using the Fama-French 3-factor model.}
	\label{tab:refinitivresultstable_FF3_INST_S5_corr}
\end{table}

\begin{table}
	\centering
	\resizebox{\textwidth}{!}{\begin{tabular}{l||c|c||c|c}
    	{} & \multicolumn{4}{c}{\textbf{Company-wise $5$-fold cross-validation}}\\
		\hline
		{} & \multicolumn{2}{c||}{Only Benchmark features} & \multicolumn{2}{c}{Benchmark and ESG features} \\
		\hline
		Year &   Balanced Accuracy & Cross-entropy loss &  Balanced Accuracy & Cross-entropy loss \\
		\hline
		2016 &      57.9 &     65.8 &     56.0 &     66.7 \\
		2017 &      55.0 &     70.6 &     55.2 &     71.6 \\
		2018 &      56.0 &     70.4 &     56.0 &     71.1 \\
		2019 &      62.4 &     64.6 &     64.7 &     64.1 \\
		2020 &      56.1 &     72.2 &     55.3 &     71.3 \\
	\end{tabular}}
	\caption{Performance measures in percent on the test set, for a target computed using the Fama-French 3-factor model.}
	\label{tab:refinitivresultstable_FF3_INST_S5}
\end{table}

\section{Relationship between validation and test cross-entropy losses for the temporal train/validation scheme}
\label{app:graphs_val_test}

\begin{figure}
	\begin{subfigure}{0.33\linewidth}
		\centering
		\includegraphics[width=1\linewidth]{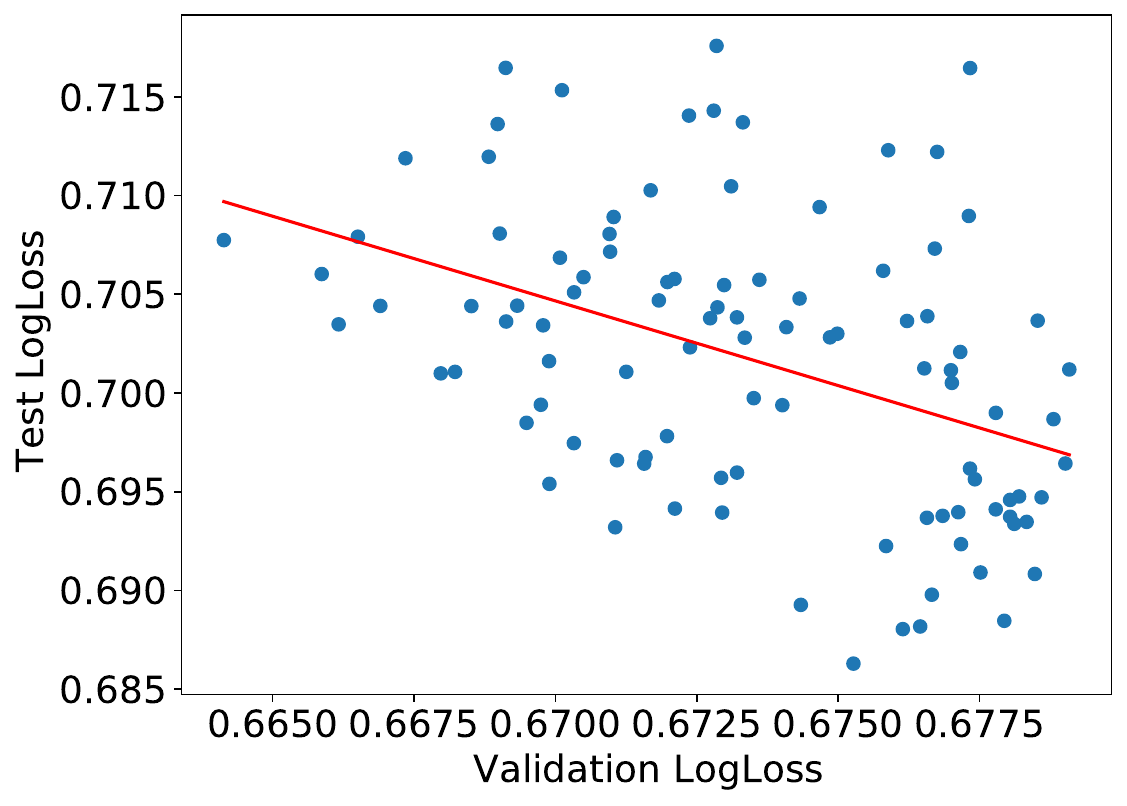} 
		\caption{2016}
		\label{fig:TEMP_42370_-0.86_0.18}
	\end{subfigure}
	\begin{subfigure}{0.33\linewidth}
		\centering
		\includegraphics[width=1\linewidth]{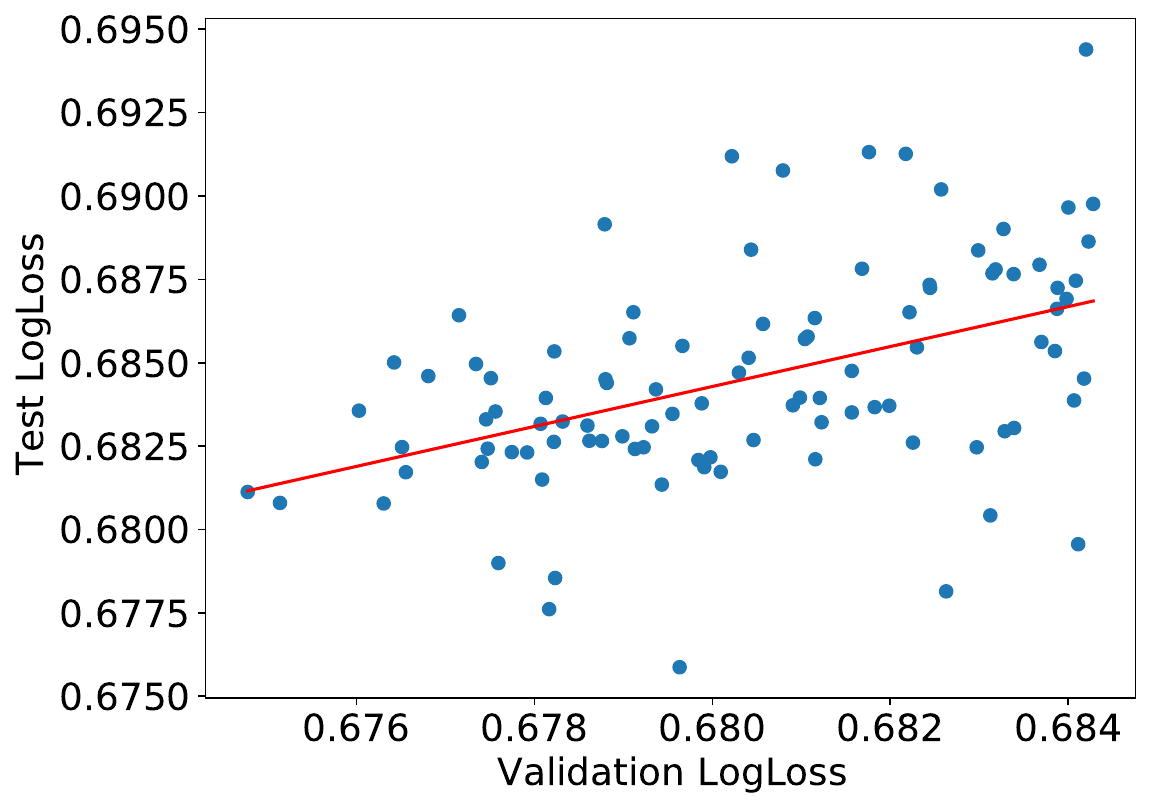}  
		\caption{2017}
		\label{fig:TEMP_42736_0.6_0.21}
	\end{subfigure}
	\begin{subfigure}{0.33\linewidth}
		\centering
		\includegraphics[width=1\linewidth]{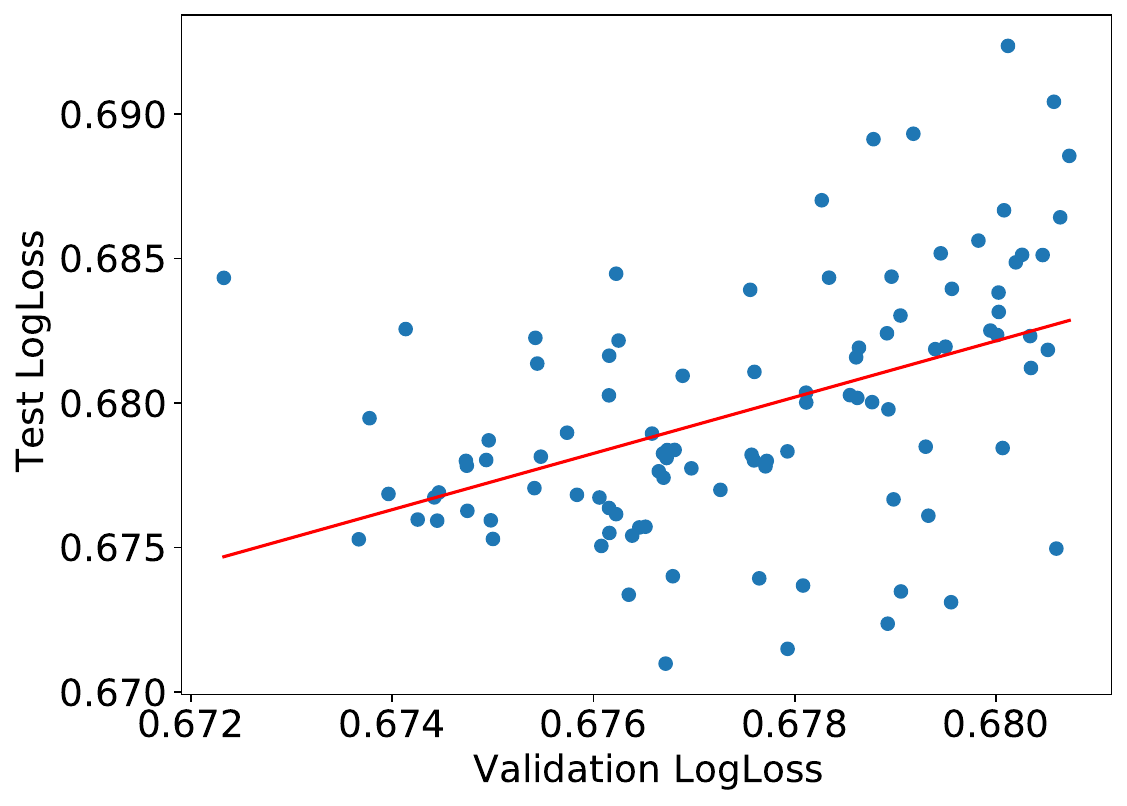}  
		\caption{2018}
		\label{fig:TEMP_43101_0.97_0.21}
	\end{subfigure}
	\begin{subfigure}{0.33\linewidth}
		\centering
		\includegraphics[width=1\linewidth]{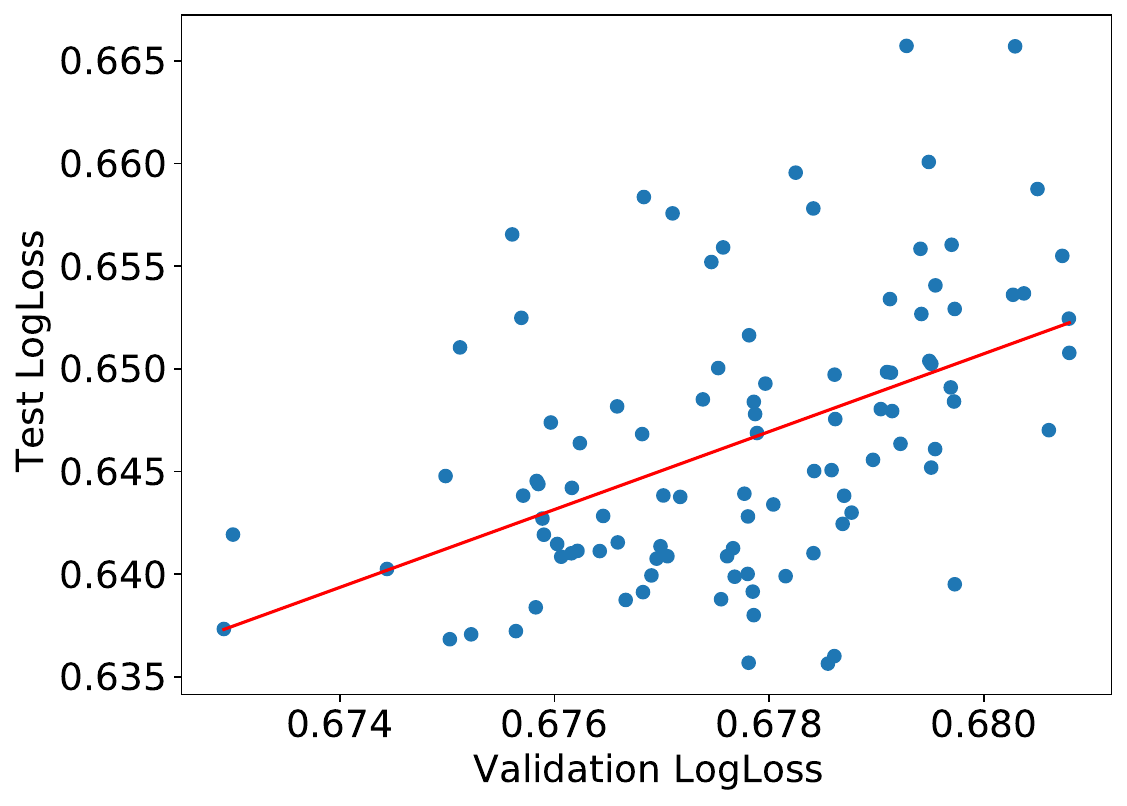}  
		\caption{2019}
		\label{fig:TEMP_43466_1.9_0.22}
	\end{subfigure}
	\begin{subfigure}{0.33\linewidth}
		\centering
		\includegraphics[width=1\linewidth]{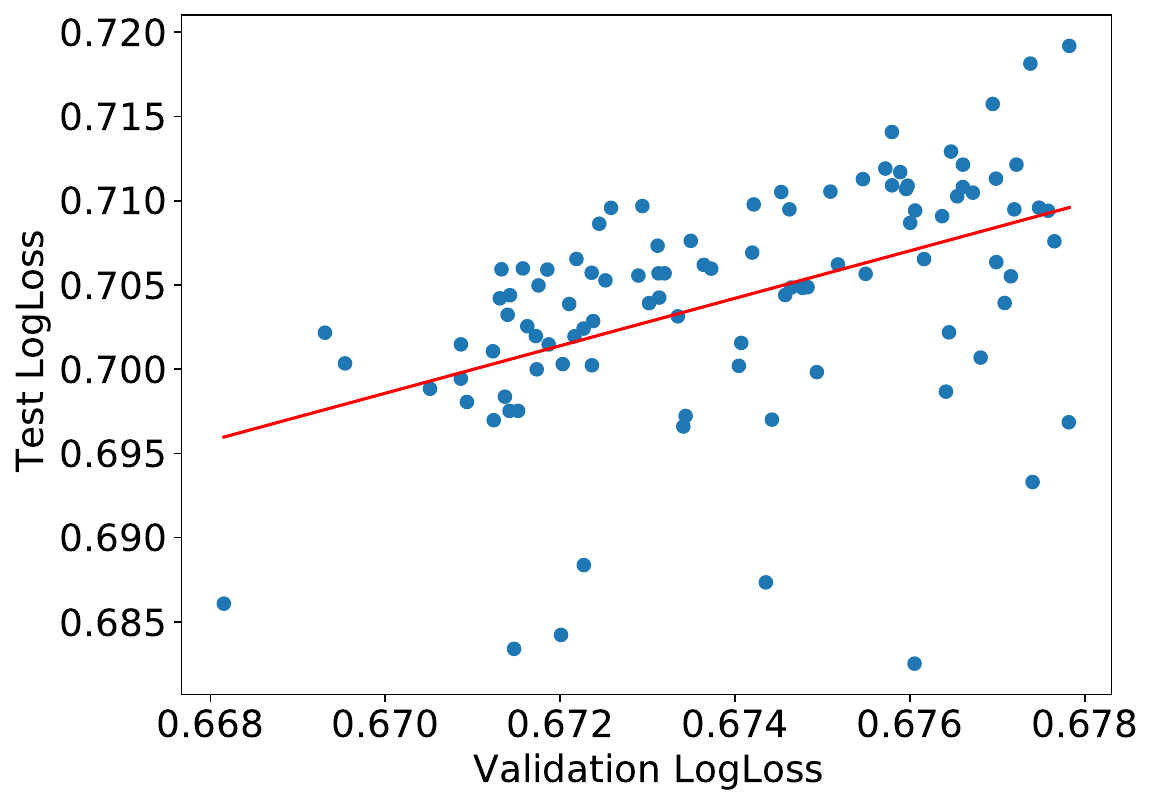}  
		\caption{2020}
		\label{fig:TEMP_43831_1.41_0.22}
	\end{subfigure}
	\caption{Standard temporal split: test set cross-entropy versus validation cross-entropy of the 100 best models of the random hyperparameters search.}
	\label{fig:refinitivresultsgraph_TEMP}
\end{figure}

We assess the relationship $\mathcal{L}_m^\textrm{test}$ versus $\mathcal{L}_m^\textrm{validation}$ for each model $m$ among the 100 models with the best validation cross-entropy losses for each of the five  sets of (train+validation)-test sets.
Figure \ref{fig:refinitivresultsgraph_TEMP} displays these relationships for the standard time-splitting scheme.

\end{document}